\documentclass[sigconf]{acmart}

\AtBeginDocument{%
  \providecommand\BibTeX{{%
    \normalfont B\kern-0.5em{\scshape i\kern-0.25em b}\kern-0.8em\TeX}}}

\makeatletter
\def\@ACM@checkaffil{
    \if@ACM@instpresent\else
    \ClassWarningNoLine{\@classname}{No institution present for an affiliation}%
    \fi
    \if@ACM@citypresent\else
    \ClassWarningNoLine{\@classname}{No city present for an affiliation}%
    \fi
    \if@ACM@countrypresent\else
        \ClassWarningNoLine{\@classname}{No country present for an affiliation}%
    \fi
}

\copyrightyear{2025}
\acmYear{2025}
\setcopyright{cc}
\setcctype{by}
\acmConference[KDD '25]{Proceedings of the 31st ACM SIGKDD Conference on Knowledge Discovery and Data Mining V.2}{August 3--7, 2025}{Toronto, ON, Canada}
\acmBooktitle{Proceedings of the 31st ACM SIGKDD Conference on Knowledge Discovery and Data Mining V.2 (KDD '25), August 3--7, 2025, Toronto, ON, Canada}
\acmDOI{10.1145/3711896.3737441}
\acmISBN{979-8-4007-1454-2/2025/08}
\usepackage[super]{nth}
\usepackage{amsmath}
\usepackage{amsthm}
\usepackage{mathtools}
\usepackage{appendix}
\usepackage{float}
\usepackage{bm}
\usepackage{dsfont}
\usepackage{graphicx}
\usepackage{caption}
\usepackage{subcaption}
\usepackage{float}
\usepackage{enumitem}
\usepackage{array}
\usepackage{makecell}
\usepackage{multirow}
\usepackage{tabularx}
\usepackage{soul}
\usepackage{url}
\usepackage{xcolor}
\usepackage[flushleft]{threeparttable}
\usepackage[ruled,vlined,linesnumbered]{algorithm2e}

\theoremstyle{definition}

\begin{document}

\title{BTS: A Comprehensive Benchmark for Tie Strength Prediction}

\author{Xueqi Cheng}
\affiliation{
  \institution{Vanderbilt University}
  \city{Nashville}
  \state{TN}
  \country{USA}
}
\email{xueqi.cheng@vanderbilt.edu}

\author{Catherine Yang}
\affiliation{
  \institution{Vanderbilt University}
  \city{Nashville}
  \state{TN}
  \country{USA}
}
\email{catherine.y.yang@vanderbilt.edu}

\author{Yuying Zhao}
\affiliation{
  \institution{Vanderbilt University}
  \city{Nashville}
  \state{TN}
  \country{USA}
}
\email{yuying.zhao@vanderbilt.edu}

\author{Yu Wang}
\affiliation{
  \institution{University of Oregon}
  \city{Eugene}
  \state{OR}
  \country{USA}
}
\email{yuwang@uoregon.edu}

\author{Hamid Karimi}
\affiliation{%
  \institution{Utah State University}
\city{Logan}
  \state{UT}
  \country{USA}
}
\email{hamid.karimi@usu.edu}

\author{Tyler Derr}
\affiliation{%
  \institution{Vanderbilt University}
  \city{Nashville}
  \state{TN}
  \country{USA}
}
\email{tyler.derr@vanderbilt.edu}

\renewcommand{\shortauthors}{Xueqi Cheng et al.}

\begin{abstract}

The rapid rise of online social networks underscores the need to understand the heterogeneous strengths of online relationships. Yet, efforts to assess tie strength (TS) are hindered by the lack of ground-truth labels, differing research perspectives, and limited model performance in real-world settings.
To address this gap, we introduce \textbf{BTS}, a comprehensive \textbf{B}enchmark for \textbf{T}ie \textbf{S}trength prediction, aiming to establish a standardized foundation for evaluating and advancing TS prediction methodologies. Specifically, our contributions are: \textbf{TS Pseudo-Label Techniques}---we categorize TS into seven standardized pseudo-labeling techniques based on prior literature; \textbf{TS Dataset Collection}---we present a representative collection of three social networks and perform data analysis by investigating the class distributions and correlations across the generated pseudo-labels; \textbf{TS Pseudo-Label Evaluation Framework}---we propose a standardized framework to evaluate the pseudo-label quality from the perspective of tie resilience; \textbf{Benchmarking}---we evaluate existing tie strength prediction model performance using the BTS dataset collection, exploring the effects of different experiment settings, models, and evaluation criteria on the results. Furthermore, we derive key insights to enhance existing methods and shed light on promising directions for future research in this domain. The BTS dataset collection, along with the curation codes and experimental scripts, is all available at: \url{https://github.com/XueqiC/Awesome-Tie-Strength-Prediction}. 
\vspace{-1ex} 
\end{abstract}

\ccsdesc[500]{Information systems~Social networks}

\keywords{Tie strength prediction; social network analysis; social tie resilience} 

\maketitle

\section{Introduction}\label{sec:intro}

Social networks have grown rapidly in recent years, becoming central to online interaction. 
A key component of these networks is the relationship between users, where the intensity of a connection---commonly referred to as \emph{tie strength (TS)}~\cite{marsden1984measuring}---has been widely studied. However, its definition can vary by context, application, or interpretation, and does not necessarily have a unified definition; for instance, a tie may be considered strong if it is bidirectional, indicating mutual interest, or may be inferred when users frequently engage with each other’s content. 
Understanding TS is critical, as it enables meaningful insights from social network data~\cite{sintos2014using}. For example, knowledge of TS can improve link prediction~\cite{li2013modeling, yang2019link}. In business networks, strong ties often correlate with successful trades and better financial outcomes~\cite{kang2021dynamic,wong2022mobile}, while in public health,  TS supports modeling disease spread, aiding prevention strategies, and healthcare planning~\cite{krendl2021impact, birditt2021age}.

During the past decades, social tie strength prediction/inference methods have been studied. The initial research developed heuristic methods to measure tie strength based on feedback from interviewees~\cite {granovetter1973strength, marsden1984measuring, petroczi2007measuring, gilbert2009predicting}. Later, researchers began to investigate using non-structural indicators such as frequency and duration of interaction~\cite{zhuang2021modeling, jones2013inferring, xiang2010modeling}, level of interaction or emotional intensity and intimacy~\cite{gilbert2009predicting,liu2020new}, and the reciprocal services found within the tie~\cite{sintos2014using, adriaens2018acquaintance, zhong2020scalable, oettershagen2022inferring}. However, these approaches generally rely on social network user characteristics that are often incomplete and difficult to measure~\cite{adriaens2018acquaintance}. To address this issue, several methods have been developed that either rely solely on the Strong Triadic Closure (STC) principle~\cite{sintos2014using, adriaens2018acquaintance, oettershagen2022inferring} or use heuristic approaches based on topological properties such as node degrees~\cite{kahanda2009using}, clustering coefficients~\cite{mattie2018understanding}, or the overlap between nodes' neighborhoods~\cite{onnela2007structure} to characterize tie strength. In recent years, methods for generating embeddings that capture latent network structures have also emerged and gained attention~\cite{zhong2020scalable, zhuang2021modeling}.

However, ground-truth tie strength label information is usually missing or incomplete in practice. Most online social network platforms lack mechanisms to distinguish between different types of ties. Furthermore, obtaining ground-truth labels of tie strengths in networks is an under-explored problem~\cite{liu2020new}. Existing literature has various tie strength definitions, including direction-based definitions~\cite{sintos2014using, adriaens2018acquaintance, zhong2020scalable, oettershagen2022inferring}, score-based definitions~\cite{liu2020new}, and frequency-based definitions~\cite{zhuang2021modeling}. 
Since the differences between these definitions vary greatly, the empirical tie strength prediction performance of these methods under different definitions also differs significantly. We therefore need standardized pseudo-label techniques to summarize these definitions to ensure consistency across studies, enabling more meaningful comparisons and allowing future research to build upon existing work more effectively.

Furthermore, as suggested by the empirical results in Section~\ref{Sec:empirical}, existing models exhibit low predictive performance and inconsistent accuracy between strong and weak ties, which can severely impact their ability to generalize across diverse real-world scenarios, limiting their effectiveness in applications such as detecting fraudulent behavior or assessing the risk of disease outbreaks within communities. Therefore, it is crucial to establish a benchmark to comprehensively evaluate and compare the performance of current and future tie strength prediction models across different settings.

To tackle these challenges, we present \textbf{BTS}, a comprehensive \textbf{B}enchmark for \textbf{T}ie \textbf{S}trength prediction, aiming to establish a standardized foundation for evaluating and advancing tie strength prediction methods. Our main contributions are as follows:
\begin{itemize}[leftmargin=*]
    \item \textbf{TS Pseudo-Label Techniques}: We categorize mainstream understandings of social tie strength into seven standardized pseudo-label definitions that assign each edge as strong or weak; 
    \item \textbf{TS Dataset Collection}: We present a well-curated collection of three social networks, offering concise representations of diverse networks that capture different real-world social interactions. We also perform dataset analysis by investigating the class distributions and correlations across the generated pseudo-labels;
    \item \textbf{TS Pseudo-Label Evaluation Framework}: 
    A standardized framework is developed to evaluate TS pseudo-labeling, assuming strong ties are more resilient than weak ties over time.
    \item \textbf{Benchmarking}: We evaluate existing tie strength prediction model performance using the BTS dataset collection, exploring the effects of different experiment settings, models, and evaluation criteria on the results. Furthermore, we derive key insights to enhance existing methods and shed light on promising directions for future research in this domain. 
\end{itemize}

The paper is structured as follows: We review existing tie strength prediction works and formalize the problem settings in Section~\ref{Sec:prelim}. We revisit mainstream understandings of tie strength and propose standardized pseudo-label techniques in Section~\ref{sec:inves}. In section~\ref{sec:dataset}, we introduce the TS dataset collection and perform empirical analysis based on it using the proposed pseudo-label techniques. In Section~\ref{Sec.resilience}, we further develop the tie resilience evaluation framework. In Section~\ref{sec:method}, we present a comprehensive benchmark of existing tie strength prediction methods. Section~\ref{Sec:lim} discusses the limitations and future work. Finally, Section~\ref{sec:con} concludes the paper.

\section{Preliminaries}~\label{Sec:prelim}
In this section, we first provide an overview of prior work focusing on measuring and predicting tie strength. Then we introduce notations and tie strength prediction settings.

\subsection{Related Work}~\label{sec:related}
With the development of online social networks, tie strength prediction has received increasing attention ~\cite{marsden2012reflections,gilbert2012predicting}. While recent papers~\cite{adriaens2018acquaintance, mattie2018understanding} have explored tie strength at fine-grained levels, the most fundamental and widely accepted understanding~\cite{granovetter1973strength, gilbert2009predicting, kumar2012using} favors the binary categorization of ties into strong/weak categories.

However, there is no standard way to obtain the ground truth tie strength~\cite{liu2020new}, and most online platforms even lack mechanisms to distinguish between different types of ties. 
To fill this gap, some of the pioneering work~\cite{petroczi2007measuring} relies on the feedback from questionnaires based on the theory proposed by ~\cite{granovetter1973strength}, that is, tie strength is a combination of the amount of time, the emotional intensity, the intimacy (mutual confiding) and reciprocal services. 

Some other works also use semantic features to infer tie strength. For example, ~\cite{xiang2010modeling} proposed a model to infer relationship strength based on profile similarity and interaction activity. However, such approaches rely heavily on the feedback qualities of users and are hard to use for large social networks. To solve the issues, recent researchers have begun to develop heuristic methods based on structural features. For example, ~\cite{mattie2018understanding} developed heuristic methods that used structure features such as degree, clustering coefficients, unweighted overlap, etc, as predictors. ~\cite{sintos2014using, adriaens2018acquaintance, zhong2020scalable, oettershagen2022inferring} also use strong triad closure properties~\cite{granovetter1973strength} to infer tie strength. Additionally, with the development of neural networks, ~\cite{liu2020new, zhuang2021modeling} developed network embedding techniques to infer tie strength. 

Despite these efforts, existing research poses different perspectives towards strong/weak ties. Some researchers~\cite{sintos2014using, adriaens2018acquaintance, zhong2020scalable, oettershagen2022inferring} built their methods based on directional-based definitions that assume bidirectional edges are strong ties, while other researchers~\cite{zhuang2021modeling, liu2020new, zhuang2021modeling, jones2013inferring, xiang2010modeling} use score or frequency criteria to determine the ground-truth tie strength. To the best of our knowledge, no previous research has investigated these understandings and how different models perform under various perspectives, which makes it challenging to ensure solid performance from existing models in different real-world scenarios. To deepen our understanding of tie strength and shed light on future directions, we explicitly conduct this paper to comprehensively study the performance of existing tie strength methods under different tie strength definitions.

\subsection{Notations}

Let $G = (\mathcal{V}, \mathcal{E}, \mathbf{X}, \mathbf{E}, \mathbf{W})$ be a directed attributed network, where $\mathcal{V} = \{v_i\}_{i = 1}^{n}$ is the set of $n$ nodes (i.e., $n = |\mathcal{V}|$) and $\mathcal{E} \subseteq \mathcal{V}\times \mathcal{V}$ is the set of $m$ edges (i.e., $m= |\mathcal{E}|$) with $e_{ij}$ denoting the edge between the node $v_i$ and $v_j$. $\mathbf{X}\in\mathbb{R}^{n\times d}$ represents the node feature matrix with dimension $d$. $\mathbf{E}\in\mathbb{R}^{m\times d'}$ represents the edge feature matrix with dimension $d'$. The edge weight matrix is $\mathbf{W}\in\mathbb{R}^{n\times n}$ where $\mathbf{W}_{ij}$ is the weight of $e_{ij} \in \mathcal{E}$. The adjacency matrix is denoted as $\mathbf{A}\in\{0, 1\}^{n\times n}$ with $\mathbf{A}_{ij} = 1$ if an observed edge exists from node $v_i$ to $v_j$ and $\mathbf{A}_{ij} = 0$ otherwise. 

\subsection{Tie Strength Prediction Problem Settings} 
In the tie strength prediction tasks, let the (pseudo) ground-truth tie strength label be $\mathbf{Y}\in\{0, 1\}^{n\times n\times C}$, where $C$ is the total number of discrete tie strength categories, and $\{\mathcal{E}_k\}_{k = 1}^{C}$ is the set of labeled ties with category $k$. Given that tie strength is a reflection of the intensity of reciprocal relationships between users, a tie is undirected, which means there is only one tie strength between node $v_i$ and node $v_j$. 
Note that in this study, we only consider binary tie strength (i.e., $C=2$) where existing edges can be either strong ($S$) or weak ($W$), which is a common setting in this line of research~\cite{liu2020new}. 
Then, the weakly-supervised and unsupervised TS prediction settings can be defined as follows:

\begin{itemize}[leftmargin=*]
    \item \textbf{Weakly-supervised settings}: Given a directed network \( G  \) with training tie strength set $\mathbf{Y}^{\text{Tr}}$, the task aims to correctly infer the strength of ties between nodes in $\mathbf{Y}^{\text{Test}}$.
    \item \textbf{Unsupervised settings}: Given a directed network \( G \) with no training tie strength set, the task aims to correctly infer the strength of ties between nodes in $\mathbf{Y}$.
\end{itemize}

\section{Tie Strength Pseudo-Label Techniques}\label{sec:inves}\label{sec:define}

In this section, we start by analyzing the existing tie strength definitions and organizing them into 3 main categories. Then, we propose a set of 7 tie strength definitions that span the primary tie strength categories from existing literature. To develop a more comprehensive understanding of these definitions, we collect three real-world datasets to conduct various analyses. Specifically, this includes analyzing the strong/weak tie distributions, pairwise correlation of strong/weak labeling of edges between definitions, and the resilience of ties from the perspective of tie dissolution.

The early work related to tie strength often relies on feedback from users who agree to participate in a survey~\cite{petroczi2007measuring, gilbert2009predicting}. However, this approach heavily depends on the quality of feedback and is hard to generalize to large-scale social networks. In recent years, researchers have developed different methods to assign labels of tie strength based on different application scenarios. Most of the efforts in this direction can be roughly categorized into three categories: 

\begin{itemize}[leftmargin=*]
    \item \textbf{Direction-based Definitions:} As suggested in ~\cite{sintos2014using, adriaens2018acquaintance, zhong2020scalable, oettershagen2022inferring}, a ``strong tie'' can be understood as a bidirectional/reciprocal relationship between users, 
    as it implies users know each other.
    \item \textbf{Score-based Definitions:} Another approach is to assign tie strength labels based on the score between users. For example, ~\cite{liu2020new} proposed assigning tie strength based on rating scores (which represent the level of trust between buyers/sellers) exchanged between anonymous users in the Bitcoin networks. 
    \item \textbf{Frequency-based Definitions:} Many works also assign tie strength labels based on frequency-related information. For instance, ~\cite{zhuang2021modeling, jones2013inferring, xiang2010modeling} all consider tie strength through the lens of how frequently users interact.
\end{itemize}

In most real-world datasets, both score-related and frequency-related information can usually be reflected in edge weight (i.e., $\mathbf{W}$). For example, the edge weight in Bitcoin\_alpha~\cite{derr2017signed} datasets represents the trust level rating score from one user to another, and the edge weight in the CollegeMSG~\cite{panzarasa2009patterns} dataset represents the contact frequency from one user to another. Hence, we summarized the existing understandings of tie strength into standardized pseudo-labels based on the edge weight and edge direction between users. More specifically, strong edges (i.e., $\mathbf{Y}_{ijS} = 1$) are assigned according to the following conditions: 
\begin{itemize}
    \item I: 
    if bidirectional between $v_i$ and $v_j$, i.e, $e_{ij}, e_{ji} \in \mathcal{E}$.
    \item II: 
    if $\mathbf{W}_{ij} + \mathbf{W}_{ji} \geq \Theta$.
    \item III: 
    if bidirectional and $\mathbf{W}_{ij} + \mathbf{W}_{ji} \geq \Theta$.
    \item IV: 
    if both $\mathbf{W}_{ij} \ge \mathbf{W}_{i}(\theta)$ and $\mathbf{W}_{ji} \ge \mathbf{W}_{j}(\theta)$.
    \item V: 
    if either $\mathbf{W}_{ij} \ge \mathbf{W}_{i}(\theta)$ or $\mathbf{W}_{ji} \ge \mathbf{W}_{j}(\theta)$.
    \item VI: 
    if $\mathbf{W}_{ij} + \mathbf{W}_{ji} \ge \mathbf{W}_{i}(\theta) + \mathbf{W}_{j}(\theta)$.
    \item VII: 
    if bidirectional and $\mathbf{W}_{ij} + \mathbf{W}_{ji} \ge \mathbf{W}_{i}(\theta) + \mathbf{W}_{j}(\theta) $.
\end{itemize}

\begin{table}[t]
\small 
\setlength\tabcolsep{4pt}
\caption{Comparison among different pseudo-labels of strong ties based on their bidirectionality and weight information.}\label{table:def_comp}
\vspace{-1ex}
\begin{tabular}{c|ccccccc}
\hline
\multirow{2}{*}{Edge Conditions}  & \multicolumn{7}{c}{Pseudo-labeling Techniques} \\ 

       & I & II & III & IV & V & VI & VII \\ \hline
Bidirectional      & \checkmark &  & \checkmark &  &  &  & \checkmark \\
Global Weight      &  & Sum & Sum &  &  &  &  \\
Local Weight      &  &  &  & And & Or & Sum & Sum \\ \hline
\end{tabular}
\vskip -1.5ex
\end{table}

In these definitions $\Theta \in (0, \infty)$ is the global threshold, and $\mathbf{W}_{i}(\theta)$ is the top $\theta$ largest edge weight in $[\mathbf{W}_{iv}, v \in \mathcal{N}_{i}]$. For each definition that requires a threshold, we first select a range of possible thresholds. For each threshold, we calculate the number of strong ties. Next, we compute the average number of strong ties across all thresholds. Finally, the threshold that produces the number of strong ties closest to this average is chosen as the final threshold. All the final thresholds can be found in the code. Based on these, we can generate the pseudo tie strength labels.

The comparisons among these pseudo-labels are shown in Table~\ref{table:def_comp}. As the semantic meaning of bi-directionality and weight information can be adjusted easily according to different datasets, we believe these 7 definitions can cover existing definitions in prior works. Moreover, the evaluations from these definitions can reveal how the models will perform in real-world scenarios.

\section{Tie Strength Dataset Collection}\label{sec:dataset}
In this study, we collect three real-world networks. We first introduce some of their details and then conduct an initial analysis. Specifically, we compare their strong/weak tie distribution and then dive into the relationship between TS definitions.

\subsection{Dataset Curation}

To cover different types of real-world social networks, we selected three datasets, including an anonymous transaction network, a college messaging network, and a popular online social network. These datasets range from public to private interactions and contain rich features for our experiments. Below are the dataset details:

\textit{Bitcoin\_alpha}~\cite{derr2017signed,arya2025datasets}: 
A cryptocurrency transaction network with anonymous online users as the nodes and their varying levels of explicitly defined trust/distrust ratings as directed weighted edges. We use text embeddings from user comments and filter out distrust edges to focus on trust-based strong/weak ties.

\textit{CollegeMSG}~\cite{panzarasa2009patterns}: This dataset is a college student interaction social network, where nodes represent students, a directed edge represents a user interacting with another user, and the edge weights represent the number of messages sent from one user to another.

\textit{Twitter}~\cite{kheiri2023analysis,farokhi2024edge}: This dataset represents a following network, where nodes are users and edges represent interactions between them (e.g., retweets and quotes). During the data collection process, we first identified users using a breadth-first search starting from a selected Twitter account, initially collecting data for 130,000 users, which includes social ties, user-generated content (such as tweets, retweets, and quotes), and, when available, user self-declared bios. To focus on active users, we only include those who have posted or deleted at least one tweet within the past three weeks. Additionally, we remove users who never interacted with others (i.e., no retweets or comments), ensuring that all remaining nodes have meaningful network connections. 
Lastly, we use the average embedding of tweets that the source user directed to the target user as the edge features, which captures the semantic context of their interaction.

\begin{table}[t]
\footnotesize
\setlength\tabcolsep{2.5pt}
\vspace{-0.75ex}
\caption{The percentage of ties labeled as strong according to various tie strength definitions for each dataset.}\label{table:strong_tie_ratio}
\begin{tabular}{c|c|c|ccccccc}
\hline
Dataset        & $n$& $m$         & I    & II   & III  & IV   & V    & VI   & VII  \\ \hline
Bitcoin\_alpha & 3,682 & 12,976  & 0.75 & 0.21 & 0.17 & 0.11 & 0.45 & 0.26 & 0.23 \\
CollegeMSG      & 1,899 & 13,838 & 0.47 & 0.24 & 0.21 & 0.11 & 0.36 & 0.21 & 0.17 \\
Twitter        & 8,351 & 175,469 & 0.24 & 0.07 & 0.04 & 0.15 & 0.27 & 0.60 & 0.13 \\ \hline
\end{tabular}
\vskip -0.5ex
\end{table}

In each of the resulting networks, we assigned an undirected tie between each pair of nodes with strength determined by the corresponding definitions. The tie weight features were obtained by averaging embeddings between two nodes for the Bitcoin\_alpha and Twitter datasets, and by taking the sum of edge weights for CollegeMSG, as no semantic edge features are available for this dataset. This transformation prepares the dataset for the experiments.

\subsection{Dataset Analysis}
\label{Sec:tie_distribution}

In this section, we apply the definitions in Section~\ref{sec:define} to the curated datasets to create the pseudo ground-truth tie strength labels. Then we present the dataset analysis to show insights related to label distribution and correlations between definitions.

\subsubsection{Label Distribution}

\begin{figure}[t]
     \centering
     \includegraphics[width=0.9\columnwidth]{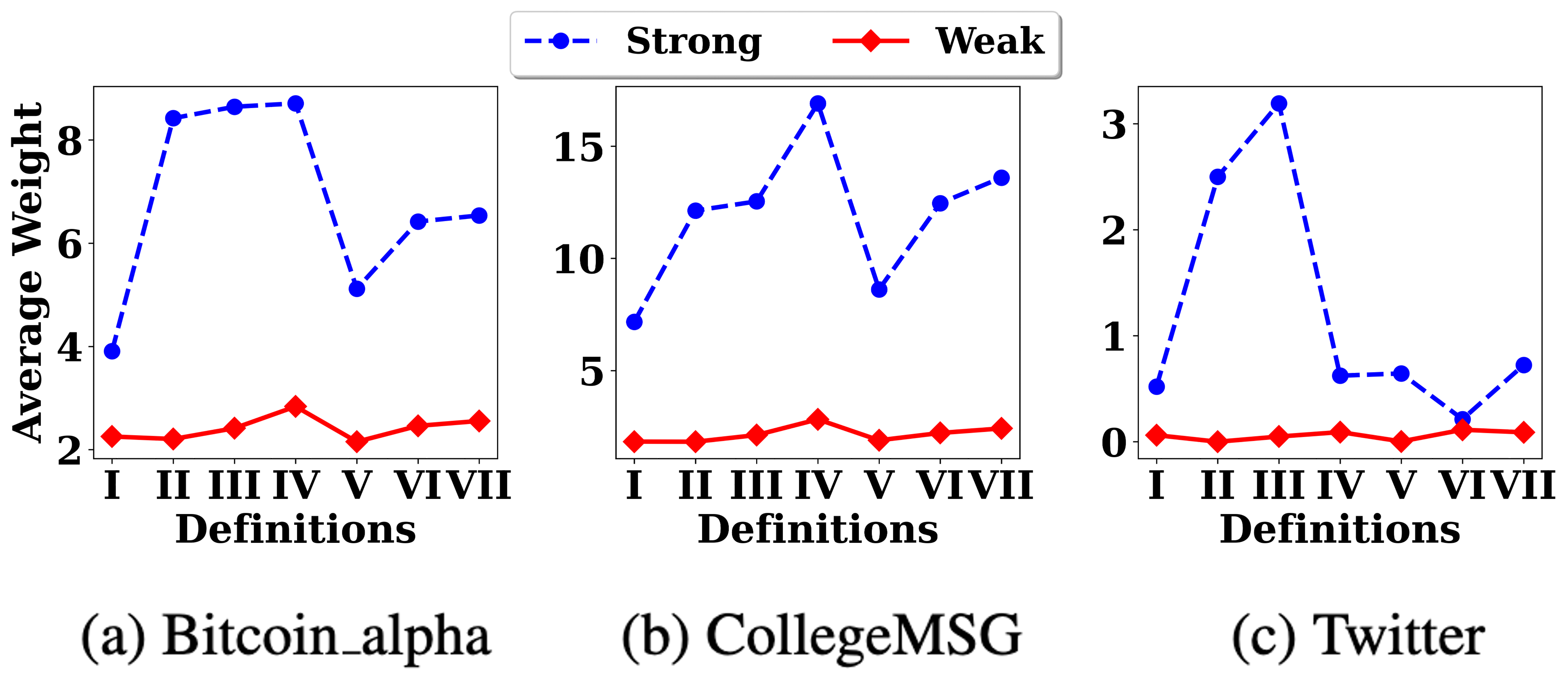}
     \vspace{-1.75ex}
     \caption{Average weights of pseudo ground-truth of strong and weak ties. The average weight of strong ties is significantly higher than the average weight of weak ties. }
     \label{fig:ave_w_0}
    \vspace{-2ex}
\end{figure}

The ratios of pseudo ground-truth strong ties under each definition for each dataset are shown in Table~\ref{table:strong_tie_ratio}. We can see that in most cases, these strong ties are the minority group in social networks. This is because strong ties usually serve as inter-community connections, and for most people, their active social circles are likely to be small in comparison with the whole network size. To further verify the effectiveness of our definitions, in Figure~\ref{fig:ave_w_0}, we show the average weight, which is calculated as the sum of the weights of all edges between the endpoint nodes of the ties for both strong and weak ties. As shown in Figure~\ref{fig:ave_w_0}, the average weight of strong ties is significantly higher than that of weak ties, demonstrating that our definitions successfully encapsulate the difference in strength between strong and weak ties.

\begin{figure}[t]
    \vspace{0.5ex}
    \centering
    \begin{subfigure}{0.33\columnwidth}
        \includegraphics[width=0.99\linewidth]{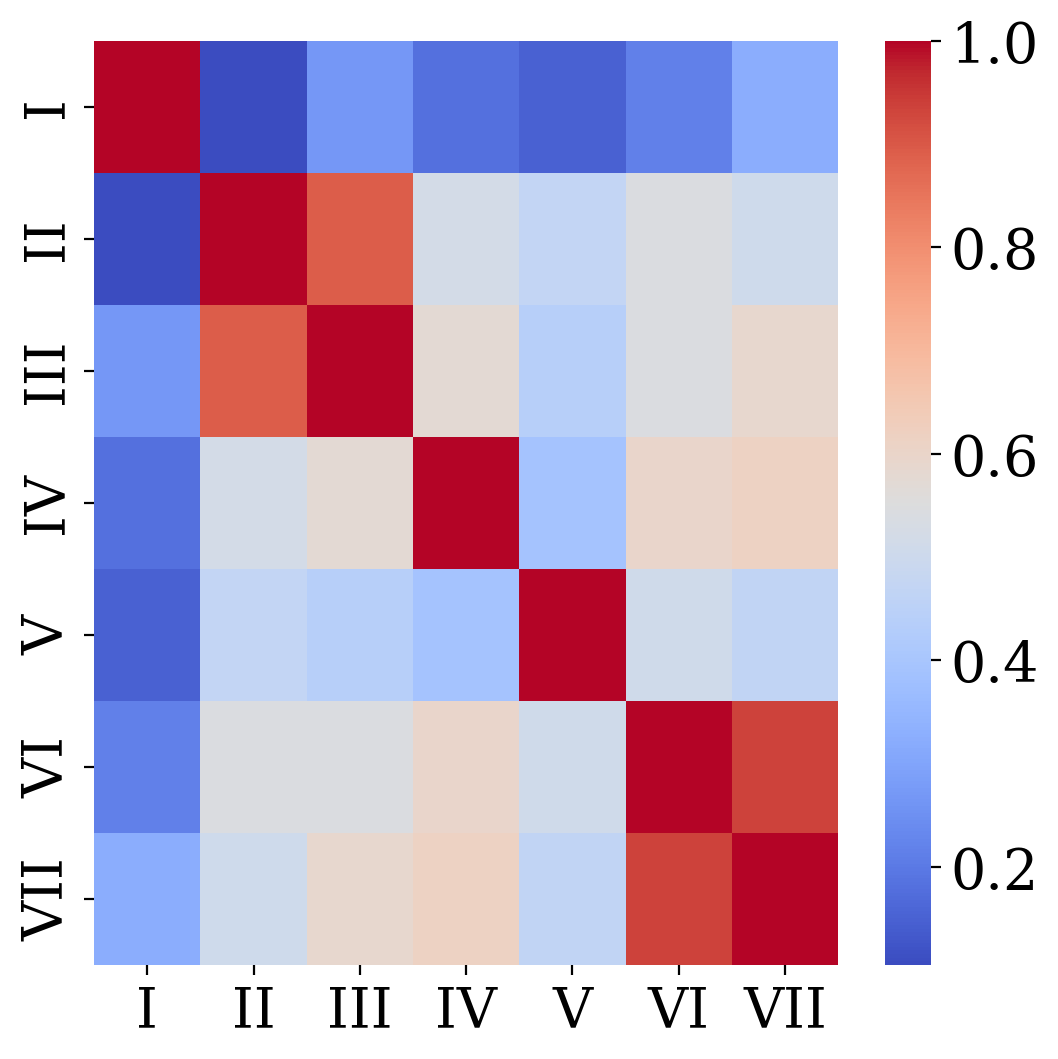}
        \label{fig:sub1}
    \end{subfigure}%
    \hfill 
    \begin{subfigure}{0.33\columnwidth}
        \includegraphics[width=0.99\linewidth]{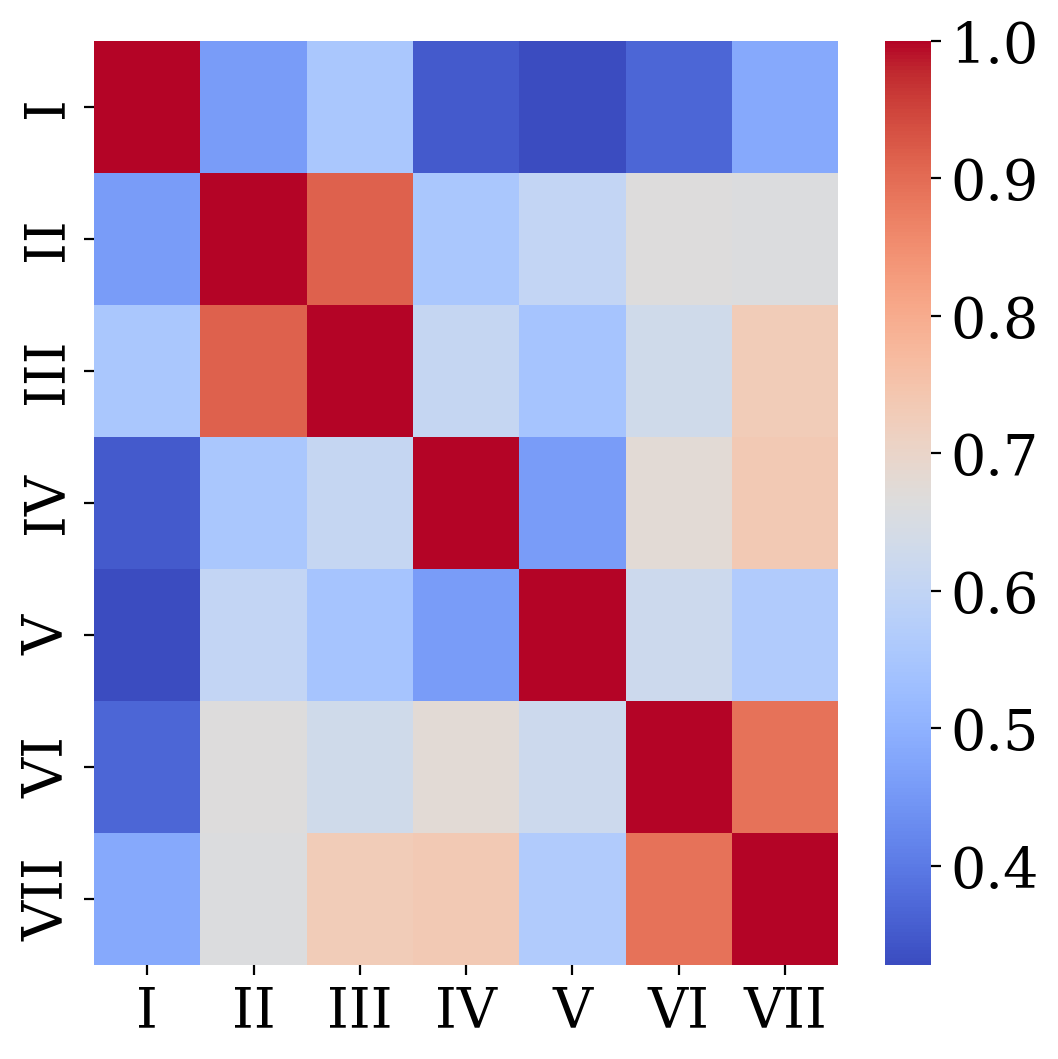}
        \label{fig:sub2}
    \end{subfigure}%
    \hfill
    \begin{subfigure}{0.33\columnwidth}
        \includegraphics[width=0.99\linewidth]{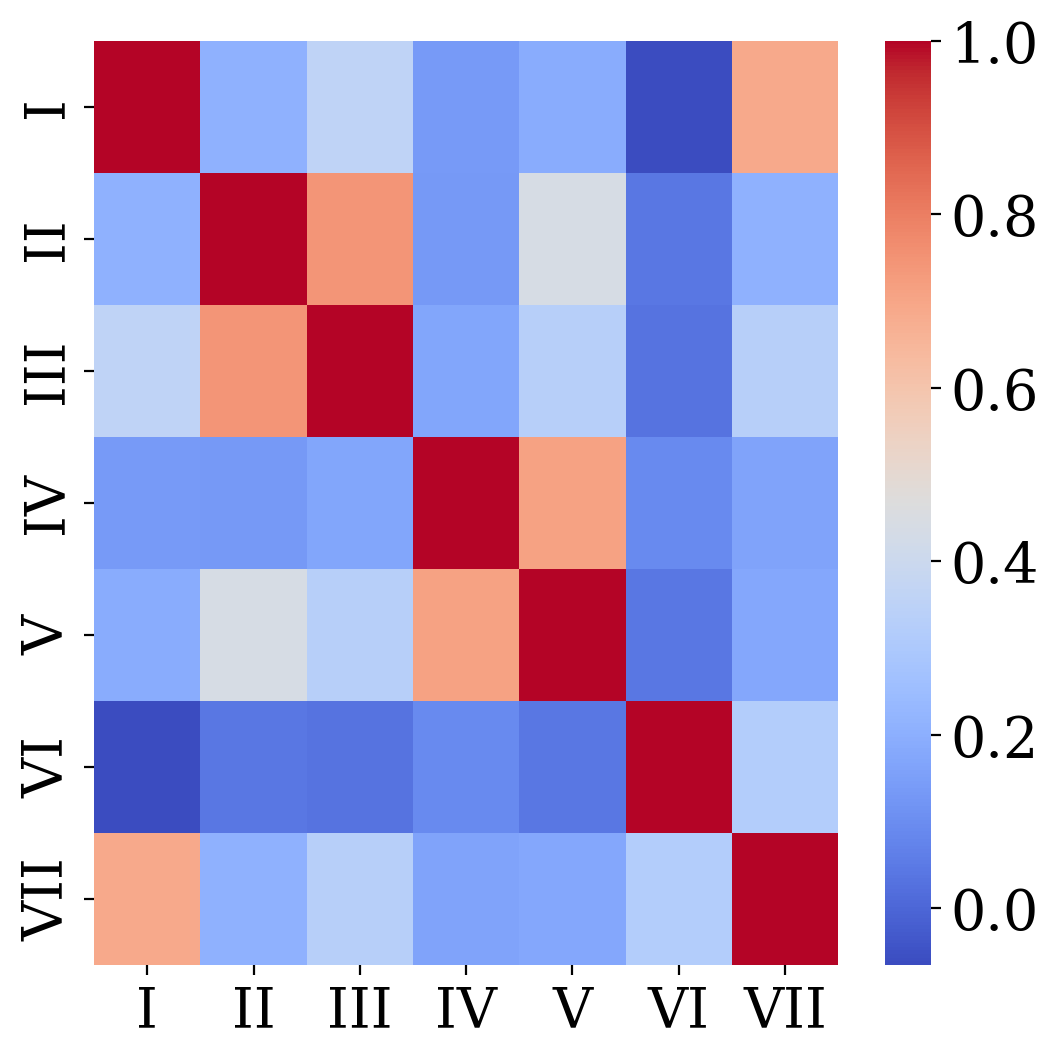}
        \label{fig:sub3}
    \end{subfigure}%
    \vskip -2.75ex
    \begin{subfigure}{0.33\columnwidth}
        \includegraphics[width=0.99\linewidth]{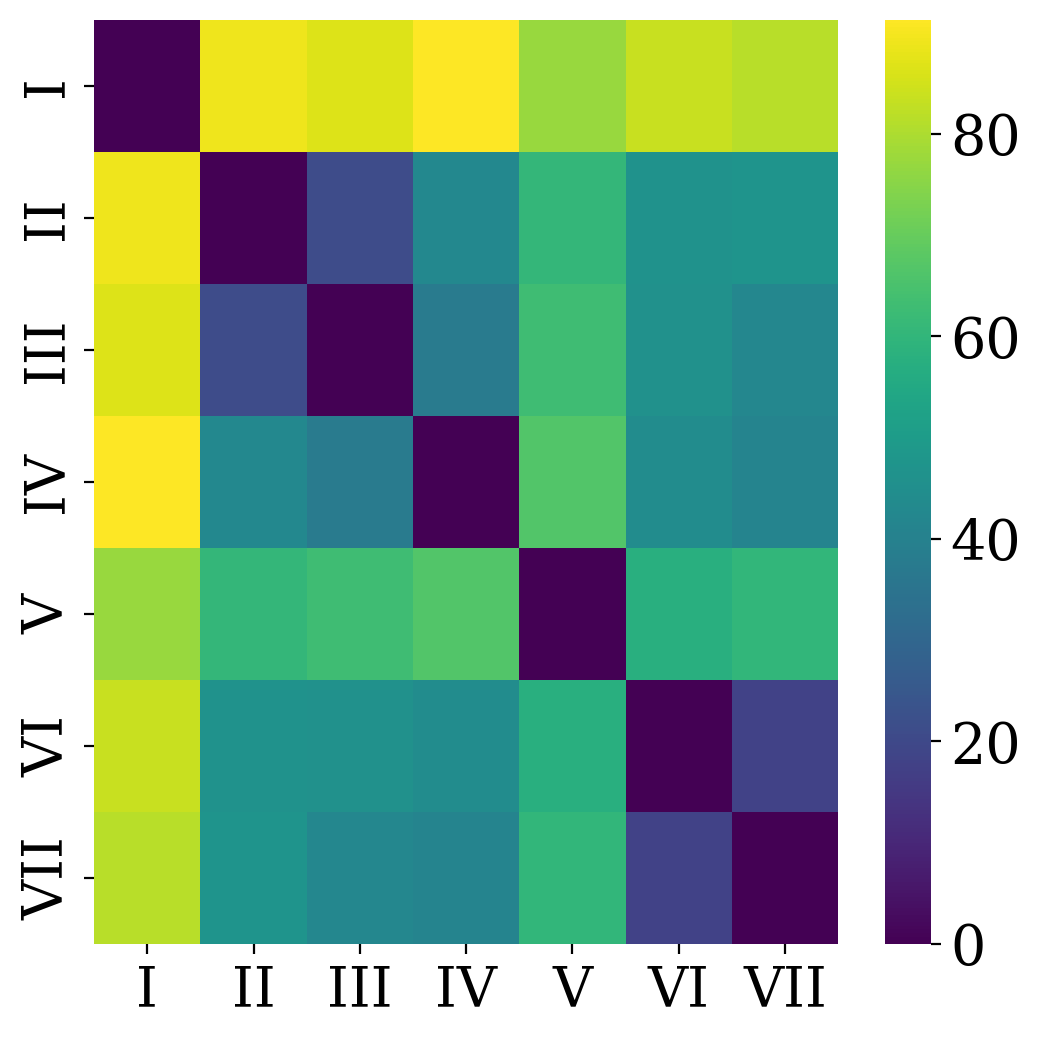}
        \caption{Bitcoin\_alpha}
        \label{fig:sub4}
    \end{subfigure}%
    \hfill
    \begin{subfigure}{0.33\columnwidth}
        \includegraphics[width=0.99\linewidth]{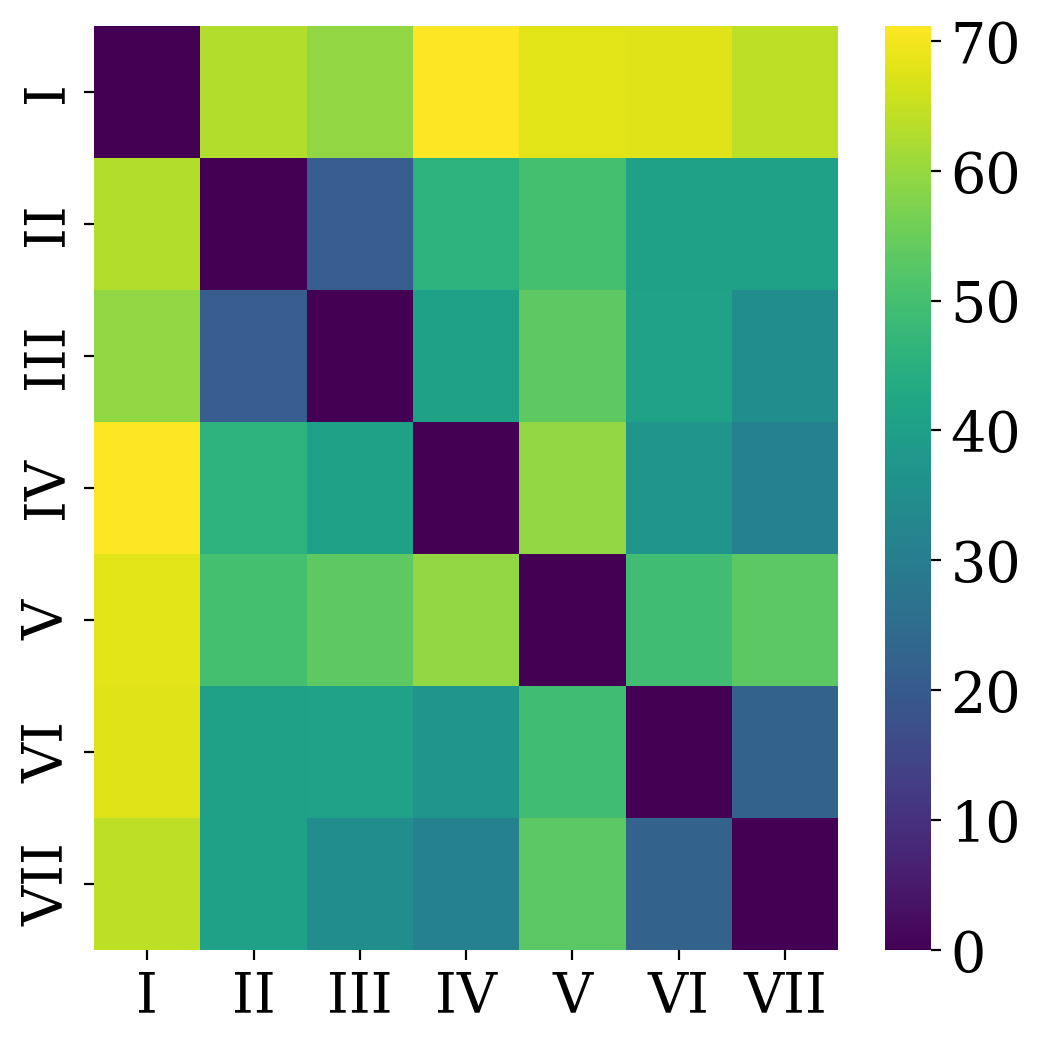}
        \caption{CollegeMSG}
        \label{fig:sub5}
    \end{subfigure}%
    \hfill
    \begin{subfigure}{0.33\columnwidth}
        \includegraphics[width=0.99\linewidth]{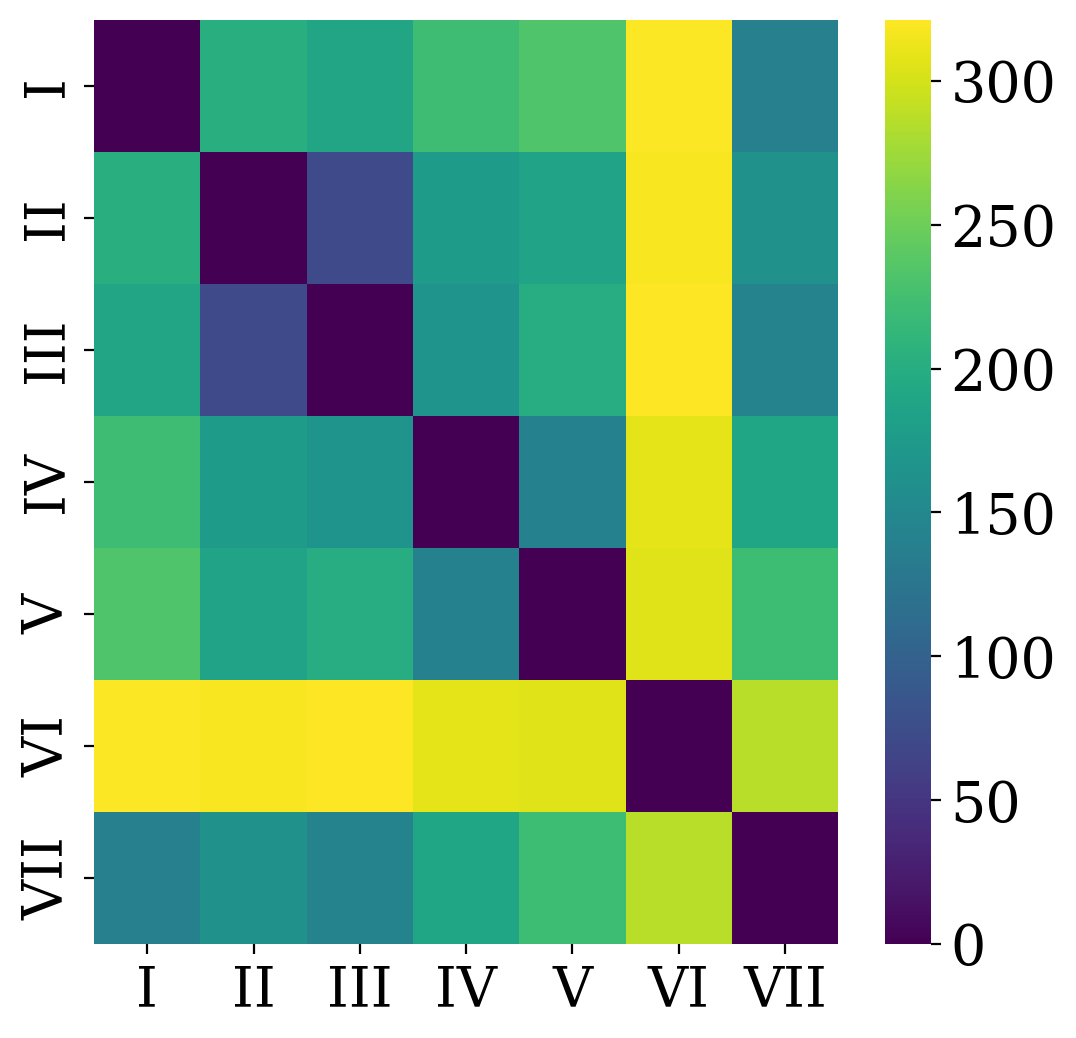}
        \caption{Twitter}
        \label{fig:sub6}
    \end{subfigure}%
    \vskip -1.25ex
    \caption{Tie strength definition relation heatmaps using Pearson correlation (top) and Euclidean distance (bottom). }\label{fig:correlation}
    \label{fig:heatmap}
    \vskip -1.5ex
\end{figure}

\subsubsection{Correlation Between Definitions}
\label{Sec.correlation}

We also use the heatmaps of Pearson correlation and Euclidean distance for these three datasets under various definitions to show their relationship. As shown in Figure~\ref{fig:heatmap}, Definition I is distinct from other definitions as it is the only one that does not consider edge weight as a criterion. Definitions II and III are more similar to each other than to other definitions because they both rely on edge weight on a global scale. Additionally, they require a tie to meet a global threshold, which is set to make sure only a small fraction of ties can achieve this in most cases. This makes such ties more likely to be bidirectional. This also stands for Definition VI and VII in our two interaction networks, Bitcoin\_alpha and CollegeMSG datasets. However, the Twitter network is a following network based on a large online social network, where one direction followings are common (as shown in Definition I in Table~\ref{table:strong_tie_ratio}). One user's frequent interaction with another does not necessarily indicate mutual following (e.g. a huge fan of some celebrities, or active users who frequently comment on a topic), but this is much more rarely the case in Bitcoin\_alpha and CollegMSG as people who don't know another are less likely to trade or message frequently. This explains why Definition VI and VII are not so similar in the Twitter network as compared with the Bitcoin\_alpha and CollegeMSG networks.

\section{Pseudo-Label Evaluation Framework}
\label{Sec.resilience}

Now that we have the pseudo-label techniques, how can we determine which ones are more appropriate for a new dataset? Inspired by the insights that stronger ties are expected to be less likely to break compared to weak ties~\cite{kivran2011impact}, we develop a novel Pseudo-Label Evaluation Framework to evaluate the quality of assigned pseudo-label from a perspective of tie resilience, where the expectation is that strong ties are more resilient and harder to dissolved (disconnected) than weak ties after time. Here we specifically define the tie resilience ratio $R$ as:
\begin{equation}\label{eq:tie_res}
    R = \frac{\text{Weak tie dissolve rate}}{\text{{Strong tie dissolve rate}}}
\end{equation} 

Then, the pseudo-label evaluation framework is proposed here: 

\begin{enumerate}
    \item \textbf{Initial Tie Strength Assignment:} 
    At time $t_0$, assign pseudo-labels to ties using the proposed method and record the counts of strong and weak ties.
    
    \item \textbf{Tracking Tie Dissolution:} After a time interval $\Delta T$, examine the set of ties originally present at $t_0$. Identify how many strong/weak ties have dissolved and calculate the strong/weak tie dissolved rates and the tie resilience ratio.
    
    \item \textbf{Repeated Observations:} Repeat the process for varying $\Delta T$ time intervals and record the results.

\end{enumerate}

\begin{table}[t]
\small 
\setlength\tabcolsep{1.5pt}
\caption{The ratio of strong/weak ties in our Twitter dataset that became dissolved after $\Delta T$ weeks under each definition. The last column/row shows the average tie resilience ratio.}\label{table:unfollow}
\begin{tabular}{c|cc|cc|cc|c}
\cline{2-8}
  & \multicolumn{2}{c|}{$\Delta T=4$} & \multicolumn{2}{c|}{$\Delta T=8$} & \multicolumn{2}{c|}{$\Delta T=12$} & \multirow{2}{*}{Avg $R$} \\ \cline{1-7}
Definition & Strong & Weak   & Strong & Weak   & Strong & Weak   &       \\ \hline
I          & 0.82\% & 0.95\% & 1.37\% & 1.78\% & 1.92\% & 2.83\% & 1.31x \\
II      & 0.77\% & 0.93\% & 1.46\% & 1.69\% & 2.18\% & 2.64\% & 1.19x \\
III        & 0.79\% & 0.92\% & 1.38\% & 1.69\% & 1.99\% & 2.63\% & 1.24x \\
IV         & 0.83\% & 0.94\% & 1.54\% & 1.70\% & 2.36\% & 2.65\% & 1.12x \\
V          & 0.89\% & 0.93\% & 1.62\% & 1.70\% & 2.42\% & 2.67\% & 1.07x \\
VI         & 0.86\% & 1.01\% & 1.51\% & 1.93\% & 2.29\% & 3.09\% & 1.27x \\
VII        & 0.83\% & 0.93\% & 1.38\% & 1.72\% & 1.81\% & 2.73\% & 1.29x \\ \hline
R & \multicolumn{2}{c|}{1.14x}        & \multicolumn{2}{c|}{1.19x}        & \multicolumn{2}{c|}{1.30x}         & 1.21x              \\ \hline
\end{tabular}
\end{table}

\begin{figure}[h]
     \centering
     \includegraphics[width=0.98\columnwidth]{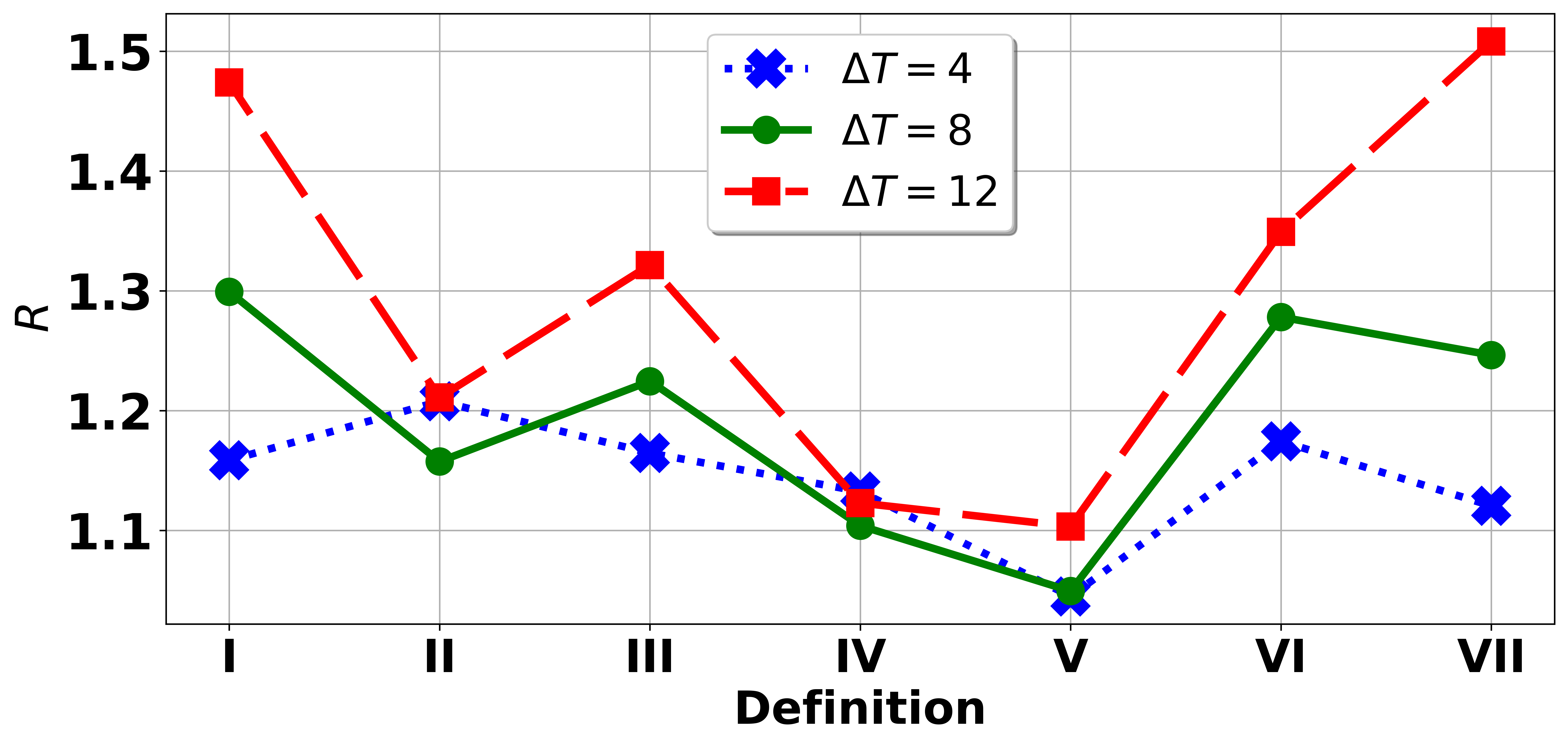}
     \vskip 0.25ex
     \caption{The dissolved rates between weak over strong ties of each definition after $\Delta T$ weeks. Overall, the ratio is becoming larger with the increase of $\Delta T$.}
     \label{fig:dis_ratio}
\end{figure}

Ideally, stronger ties should have a lower dissolution rate than weaker ties ($R > 1$), and the tie resilience ratio is expected to increase as the network evolves. To examine this, we analyze the Twitter dataset, with results presented in Table~\ref{table:unfollow}. Overall, strong ties dissolve less frequently than weak ties across all definitions, and both types of ties become more likely to dissolve as $\Delta T$ increases. More specifically, Definitions I, III, and VII, which define strong ties as bidirectional connections, exhibit lower disconnection rates, reinforcing the idea that bidirectional ties are more stable than unidirectional ones. Additionally, as $\Delta T$ increases, the probability of both strong and weak ties breaking rises. Figure~\ref{fig:dis_ratio} further confirms this trend, showing that $R$ increases with $\Delta T$, indicating that weak ties deteriorate faster than strong ties. These findings support our intuition and validate the correctness of the evaluation framework.

Based on the results in Table~\ref{table:unfollow}, we gain key insights into the quality of pseudo-labels assigned under different definitions. For instance, Definition I (Avg. $R=1.31$) provides a more meaningful representation of tie strength than Definition V (Avg. $R=1.07$). This aligns with our understanding of Twitter’s social structure, where Definition I (users following each other) better represents a strong tie compared to Definition V (one user frequently commenting on another), which may simply reflect a highly engaged follower. This analysis demonstrates that our framework effectively evaluates the quality of pseudo-labels based on the dataset's characteristics.

\section{Benchmark for Tie Strength Prediction}\label{sec:method} 

Now, after having defined the proposed tie strength pseudo-labeling techniques, curated dataset collection, and pseudo-label evaluation framework, in this section, we introduce the details of our benchmark for tie strength prediction. Specifically, we present details on multiple real-world TS prediction settings, representative models used for later experiments, and evaluation criteria for the tie strength prediction benchmarking. 

\subsection{Experiment Settings}~\label{sec:exp_setting}
Based on Section~\ref{sec:define}, we can now assign pseudo ground-truth tie strength labels in networks. However, as discussed in Section~\ref{sec:intro}, the ground-truth tie strength information is usually incomplete or missing, and most online platforms lack mechanisms to distinguish between different types of ties. Therefore, when evaluating the existing tie strength prediction models' performance, the common supervised learning setting with many labeled edges may not align well. A more realistic setting would be weakly supervised, which stands for the ground-truth label is incomplete, or even unsupervised learning, where all ground-truth information is missing. Moreover, compared with the traditional random weakly supervised setting, a node-centric training selection might better align with the real-world social science domain; i.e., the idea of conducting surveys from a few individuals who provide their tie strengths (compared to randomly distributed user pairs). Based on these thoughts, we propose two experimental settings:

\begin{itemize}[leftmargin=*]

\item \textbf{Random weakly supervised setting:} Randomly splitting the training edge set $\mathcal{E}^{\text{Tr}}$, validation edge set $\mathcal{E}^{\text{Val}}$, and testing edge set $\mathcal{E}^{\text{Test}}$, where $|\mathcal{E}^{\text{Tr}}| = |\mathcal{E}^{\text{Val}}| << |\mathcal{E}^{\text{Test}}|$. 
    \vspace{0.5ex}

\item \textbf{Node-centric weakly supervised setting:} Randomly select a small number of nodes as hub nodes to form the training hub node set $\mathcal{V}^{\text{Tr}}_{\text{hub}}$ ($|\mathcal{V}^{\text{Tr}}_{\text{hub}}| << |\mathcal{V}|$). Then randomly select the ties connected with these nodes to form the training edge set $\mathcal{E}^{\text{Tr}}$. In the remaining network, randomly select validation hub nodes set $\mathcal{V}^{\text{Val}}_{\text{hub}}$ such that $|\mathcal{V}^{\text{Val}}_{\text{hub}}| = |\mathcal{V}^{\text{Tr}}_{\text{hub}}|$, and get the validation edge set $\mathcal{E}^{\text{Val}}$ same as above. All the remaining edges in the network form the testing edge set $\mathcal{E}^{\text{Test}}$.
    \vspace{0.5ex}

\item \textbf{Unsupervised setting:} In this setting, unsupervised models utilize the entire network without access to any pseudo ground-truth information regarding tie strength. The inferred tie strengths are then compared against the pseudo ground-truth labels to evaluate the models' performance across various definitions.
    \vspace{0.5ex}

\end{itemize}

In this study, the weakly supervised setting is transductive, and all edges in the training set are considered labeled. Inspired by the N-way K-shot setting in few-shot learning~\cite{cheng2019few}, we explicitly set the size of the validation set to be equal to the size of the training set. To make sure the results between weakly supervised settings I and II are comparable, we explicitly set the number of training edges of these two settings to be approximately the same. Specifically, for the random weakly supervised setting, the training edge set sizes are 20 for Bitcoin\_alpha, 70 for CollegeMSG, and 50 for Twitter. For the node-centric weakly supervised setting, the number of hub nodes is set to 5 for Bitcoin\_alpha, 4 for CollegeMSG, and 2 for Twitter, resulting in an average of 21, 68.3, and 52.3 training edges, respectively, across three splits. 

\subsection{Models for Tie Strength Prediction}
As in Section~\ref{sec:related}, this study considers three main approaches to infer tie strength: heuristic methods, STC-based methods, and neural network methods. We focus on using structural and edge features.

For heuristic methods, we adopted the approach from ~\cite{mattie2018understanding} by training a Random Forest model~\cite{breiman2001random} using structural features such as common neighbors, the sum of end nodes' degrees, the sum of clustering coefficients, etc. For STC-based methods, we used STC-Greedy code from~\cite{adriaens2018acquaintance}. For neural network methods, inspired by ~\cite{zhuang2021modeling}, we selected two variants of a 3-layer Multi-layer Perception (MLP)~\cite{taud2018multilayer}, one that includes edge features and one that does not, and two Graph Neural Networks (GNNs): GCN~\cite{kipf2016semi} and GTN~\cite{shi2020masked}, where one key difference is whether edge features are utilized in the message-passing process.

\subsection{Evaluation Criteria}

Following the evaluation setting of~\cite{gilbert2009predicting, sohrabi2016comprehensive}, we adopt accuracy as the primary evaluation metric. Additionally, to better quantify the disparity in intensity between strong and weak ties, similar to~\cite{sintos2014using, adriaens2018acquaintance}, we also use the average tie weight difference as an additional criterion to evaluate the prediction quality.

\section{Empirical Study of Tie Strength Prediction}\label{Sec:empirical}

In this section, we present the results of existing models and offer an in-depth discussion. All results are averaged over 3 runs with different data splits. The experiments were performed on a desktop with a 24GB NVIDIA GeForce RTX 3090 GPU and 64GB RAM. 
Our code and datasets are all publicly available\footnote{\url{https://github.com/XueqiC/Awesome-Tie-Strength-Prediction}}.

\subsection{Random Weakly Supervised Setting}\label{Sec:emp_weak1}

\begin{figure}[t]
    \centering
    \includegraphics[width=0.85\columnwidth]{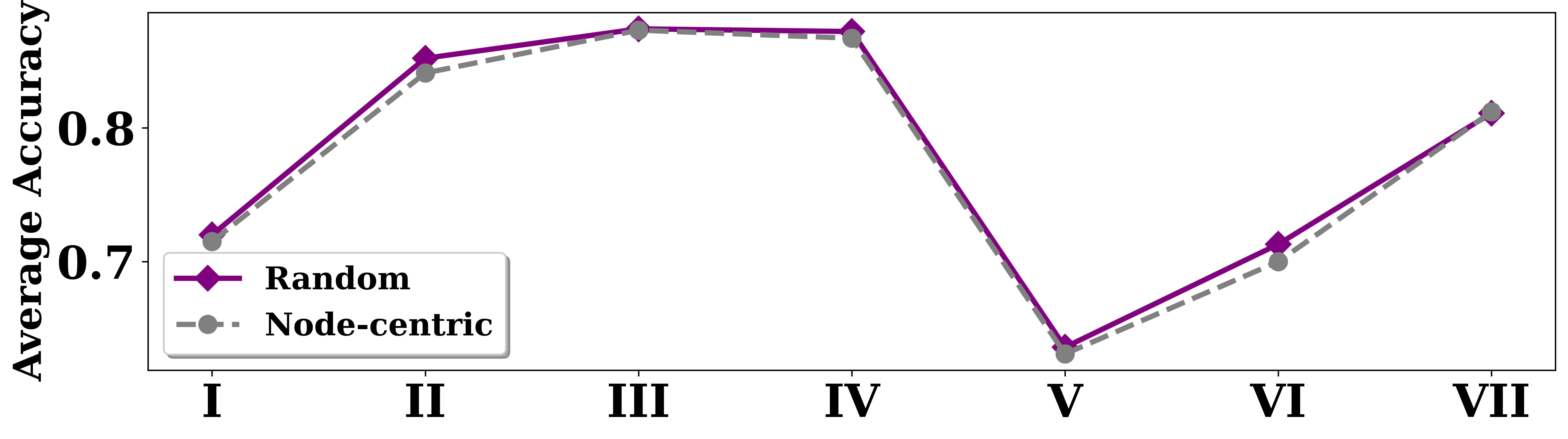}
    \vspace{-2ex}
    \caption{Average prediction accuracy of two weakly supervised settings across all models and datasets.} 
    \label{fig:settingvs}
    \vspace{-2ex}
\end{figure}

\begin{itemize}[leftmargin=*]
    \item \textbf{Nerual network models perform better}: As shown in Table~\ref{table:res1} (in the Appendix), neural network methods outperform heuristic methods in prediction accuracy, with MLP w/ $\mathbf{E}$ achieving the best performance in the Bitcoin\_alpha dataset, and GCN performing the best in CollegeMSG and Twitter datasets. In contrast, heuristic methods depict the worst overall performance in Bitcoin\_alpha and CollegeMSG, and the second worst in Twitter. This highlights the greater overall effectiveness of neural network methods in leveraging structural and semantic features to predict tie strength, and heuristic methods that are built on pre-established rules may not be easily adaptable to different real-world scenarios. The fact that MLP w/ $\mathbf{E}$ depicts better than MLP w/o $\mathbf{E}$, but GCN performs better than GTN shows that leveraging semantic features has the potential to improve the performance, but GNNs need better mechanisms to integrate edge features.
    \vspace{0.5ex}

    \item \textbf{Models struggle under certain pseudo-labels techniques}: 
    When analyzing the results from Figure~\ref{fig:settingvs} across different definitions, we observe that the existing methods vary significantly in their performance. Specifically, Definitions I, V, and VI tend to have lower prediction accuracy. To analyze the reasons for these performance drops, we review the properties related to these definitions and find that these definitions are the most distinct cases according to Figure~\ref{fig:correlation}. More specifically, when looking into Definition I, we can find that performance is poor in Bitcoin\_alpha and CollegeMSG but better in Twitter. This is likely because in Bitcoin\_alpha and CollegeMSG, Definition I is less correlated with other definitions compared to Twitter. A similar pattern is seen in Definitions V and VI. This suggests that 1) there may be limitations in these techniques that cause them to assign pseudo ground-truth labels in a way that differs significantly from other definitions, and 2) model performance is highly dependent on the pseudo-labels techniques used. This again highlights the importance of ensuring consistency across studies so that more meaningful comparisons can be made.
    \vspace{0.5ex}

    \item \textbf{Imbalance issue}: In definitions with the highest overall performance, such as Definition III and IV, there are still concerns related to cases of extreme imbalance, like Twitter in Definition III and CollegeMSG in Definition IV. Based on the existing literature related to imbalance, the seemingly high prediction performance can be ``unreal" as skewed distribution favors predictions toward the majority class, typically weak ties, which negatively impacts prediction accuracy for the minority class. We explore this imbalance issue further in Section~\ref{sec:further}.
    \vspace{0.5ex}

    \item \textbf{Insignificant average tie weight differences}: The weight of a tie is the sum of all edge weights between its end nodes. Average weight differences between strong/weak ties across different datasets are shown in Table~\ref{table:ave_ba1}. As edge weight reflects the intensity of interactions or the level of trust, we expect the models can successfully reflect this by assigning strong labels to ties with relatively high edge weight. However, compared with the pseudo ground-truth, the existing methods fail to depict significant average weight differences across all definitions, which implies that the existing methods fail to capture the assumed differences in strong versus weak ties having higher and lower edge weights, respectively. This could be because the models suffer from the imbalance issues, and/or are affected by noises in features and local structure information. 
\end{itemize}

\vspace{-2ex}

\subsection{Node-centric Weakly Supervised Setting}\label{Sec:emp_weak2}

The prediction accuracy for node-centric random weakly supervised settings is presented in Table~\ref{table:res2} (in the Appendix). Similar to the results in Section~\ref{Sec:emp_weak1}, neural network methods outperform heuristic methods in Bitcoin\_alpha and CollegeMSG, with MLP w/ $\mathbf{E}$ achieving the best performance, while heuristic methods perform poorly in these datasets. However, heuristic methods show the best overall performance on Twitter, mainly due to their strong performance in Definition VI, where the predictors employed by heuristic methods successfully distinguish strong and weak ties. The poorer performance of neural network models here compared to Table~\ref{table:res1} is likely because these models rely on hub nodes' features and local class distribution information, which can introduce bias when this local information deviates significantly from global contexts.

In terms of performance across different definitions, we observe lower accuracy in Definitions I, V, and VI, consistent with Section~\ref{Sec:emp_weak1}. This reinforces that model performance depends on the tie strength definition used; thus, multiple definitions should be used to ensure a robust benchmarking. We also observe that models exhibit biased performance under extreme imbalance, such as Twitter (Definition III) and CollegeMSG (Definition IV). Additionally, average tie weight differences across datasets (Table~\ref{table:ave_ba2}) are less pronounced than those in the pseudo ground-truth; further discussed in Section~\ref{sec:further}. Lastly, we emphasize the need for a node-centric weakly supervised setting beyond the traditional random setting when empirically benchmarking tie strength prediction methods, as it aligns with many real-world scenarios (as discussed in Section~\label{sec:exp_setting}).

\begin{figure}[t]
    \centering
    \begin{subfigure}{0.45\columnwidth}
        \includegraphics[width=\linewidth]{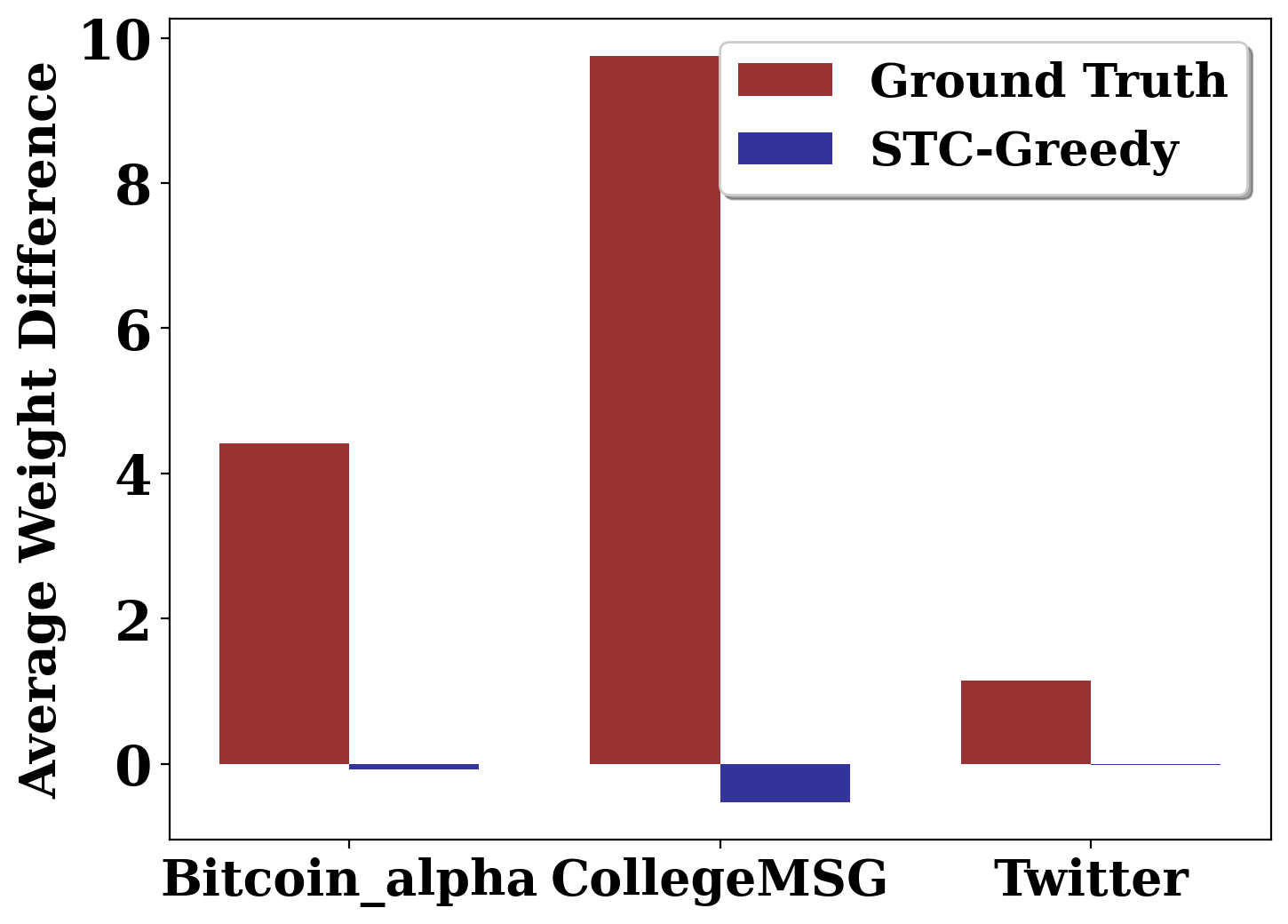}
        \caption{Datasets}
        \label{fig:ave_weight3}
    \end{subfigure}%
    \hfill
    \begin{subfigure}{0.45\columnwidth}
        \includegraphics[width=\linewidth]{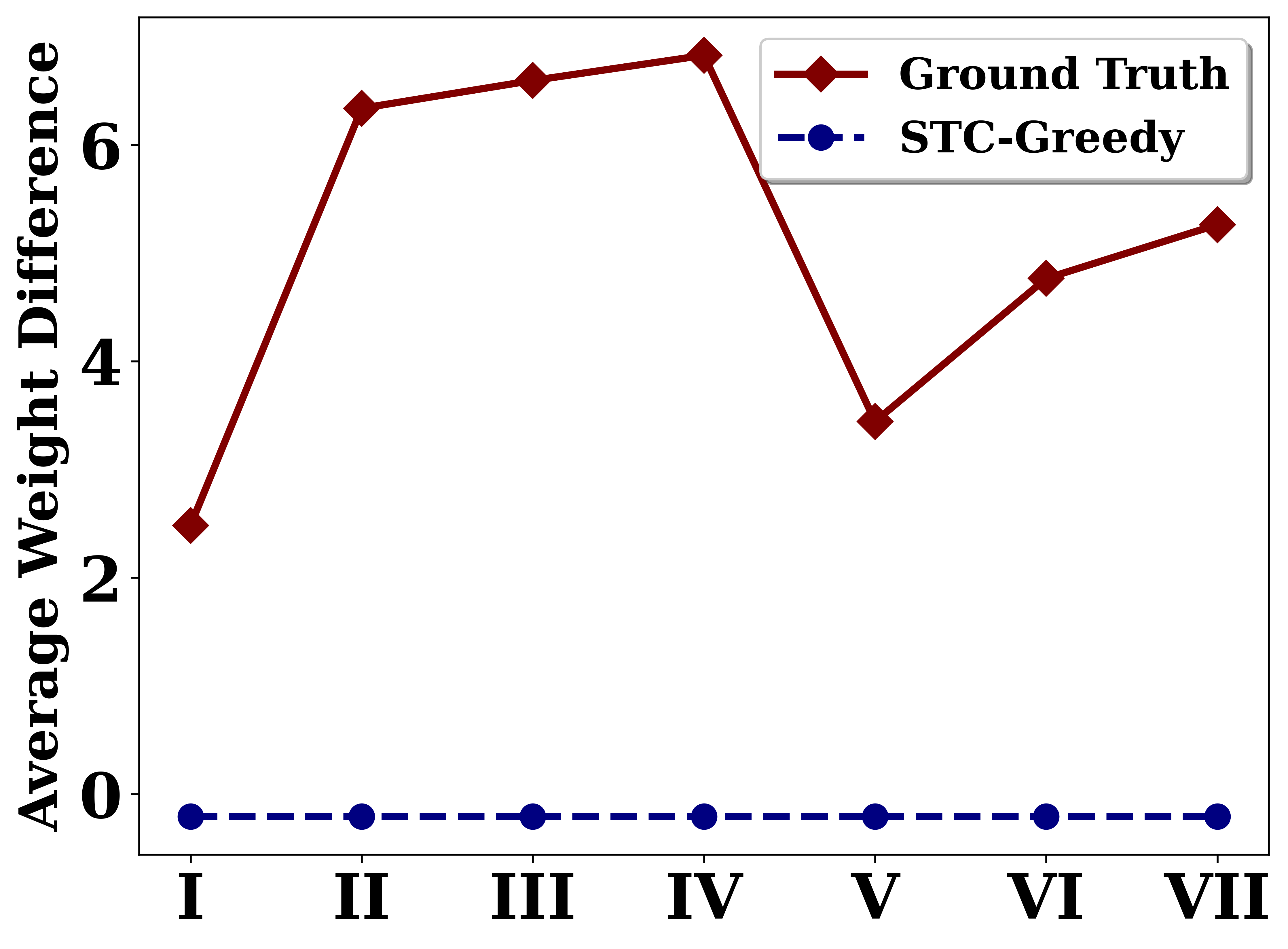}
        \caption{Definitions} 
        \label{fig:ave_weight31}
    \end{subfigure}%
    \vskip -1.5ex
    \caption{Average weight differences of applying STC-Greedy compared with ground truth across different datasets and definitions. Overall, STC-Greedy depicts low performance and is not applicable for complex situations}\label{fig:ave_stc}
    \vspace{-3ex}
\end{figure}

\subsection{Unsupervised Setting}\label{Sec:emp_un}

The inference accuracy of using STC-Greedy is presented in Table~\ref{table:res3}. Since no label information is provided for training in this setting, the accuracy of STC-Greedy is understandably lower compared to the prediction accuracy discussed in the previous two sections. However, we still observe a notable performance drop in Definitions I, V, and VI, particularly in Bitcoin\_alpha and CollegeMSG for Definitions I and V, and in Twitter for Definition VI. This aligns with our previous observations and further demonstrates that model performance is highly dependent on the definition used.

The results for the average weight differences using STC-Greedy across different datasets (Figure~\ref{fig:ave_weight3}) and definitions (Figure~\ref{fig:ave_weight31}) show consistently insignificant differences compared to the ground truth. Note that since STC-Greedy cannot learn from the pseudo ground truth label from different definitions, their inference results are not changing. This also indicates that relying solely on STC properties cannot sufficiently highlight edges with larger weights, and better models are necessary to achieve improved performance.

\begin{figure*}[h!]
    \hspace{1ex}\begin{subfigure}{0.6\columnwidth}
        \includegraphics[width=\linewidth]{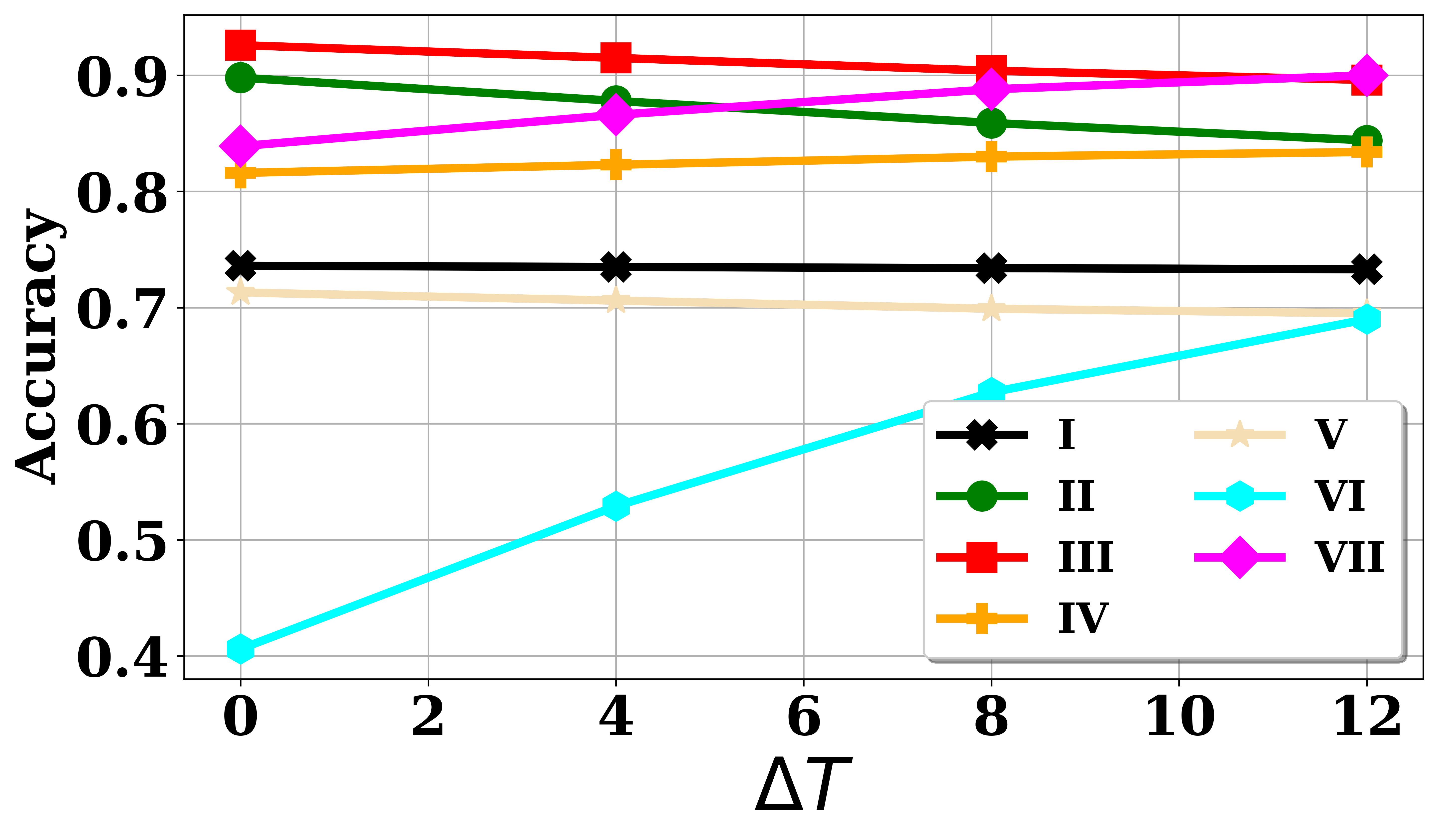}
        \vskip -0.75ex\caption{STC-Greedy}
        \label{fig:stc_future}
    \end{subfigure}%
    \hfill
    \begin{subfigure}{0.6\columnwidth}
        \includegraphics[width=\linewidth]{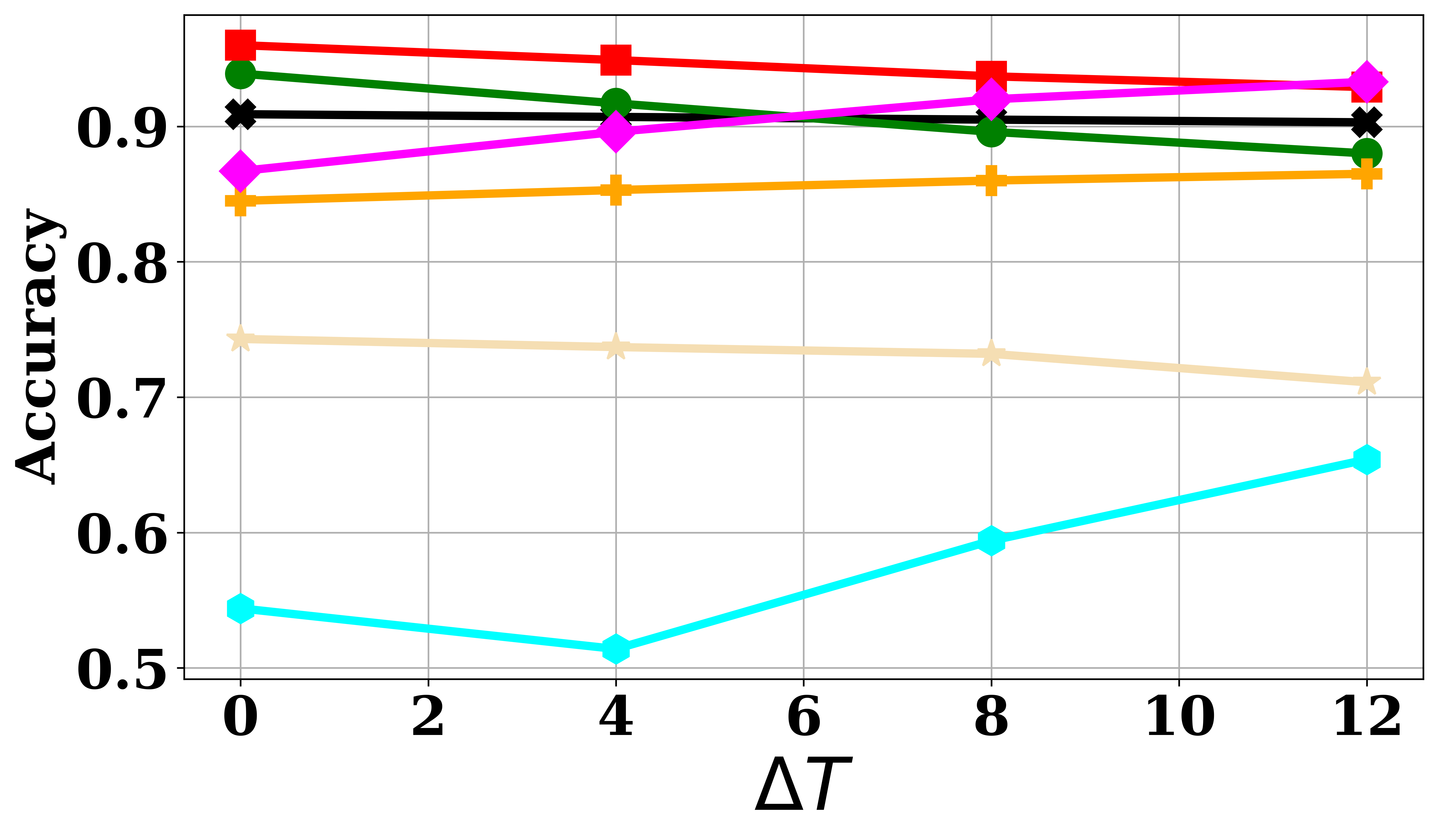}
        \vskip -0.75ex\caption{MLP w/ $\mathbf{E}$}
        \label{fig:mlp_future}
    \end{subfigure}%
    \hfill
    \begin{subfigure}{0.6\columnwidth}
        \includegraphics[width=\linewidth]{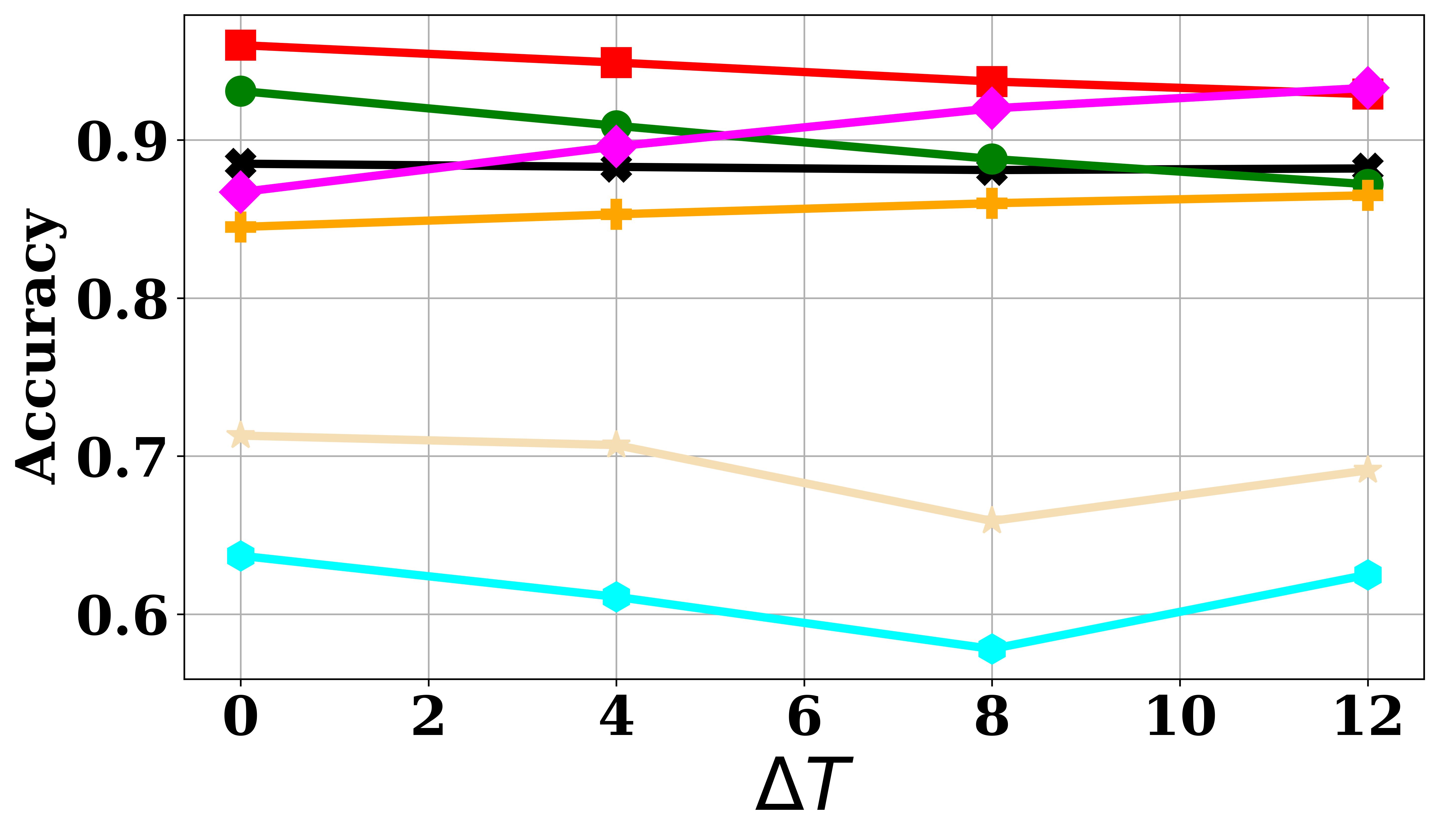}
        \vskip -0.75ex\caption{GCN}
        \label{fig:gcn_future}
    \end{subfigure}%
    \hspace{1ex}
    \hfill 
    \vskip -1ex
    \caption{Prediction accuracy of predicting tie strength in the Twitter dataset after $\Delta T$ weeks, based on the initial network.} \label{fig:pred_future}
     \vskip -2ex
\end{figure*}

\begin{figure*}[h!]
    {\centering
    \begin{subfigure}{2\columnwidth}
     \centering
        \includegraphics[width=0.75\linewidth]{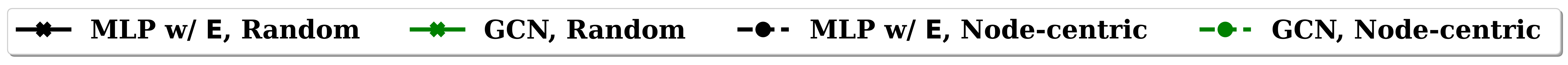}
    \end{subfigure}%
    
    \begin{subfigure}{0.66\columnwidth}
      \centering
        \includegraphics[width=0.8\linewidth]{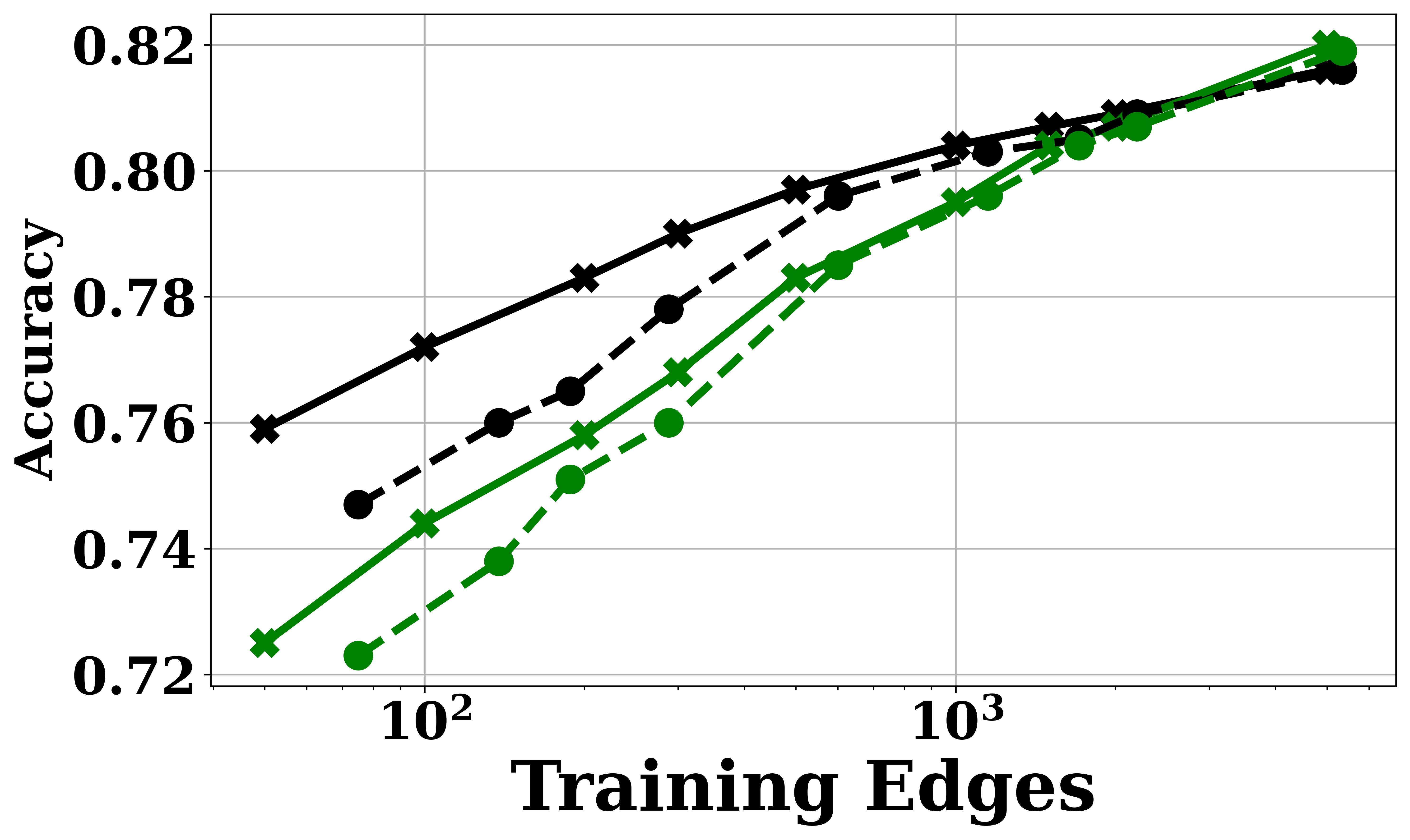}
        \vskip 0.25ex\caption{Bitcoin\_alpha}
        \label{fig:ba_labels}
    \end{subfigure}%
    \hfill
    \begin{subfigure}{0.66\columnwidth}
      \centering
        \includegraphics[width=0.8\linewidth]{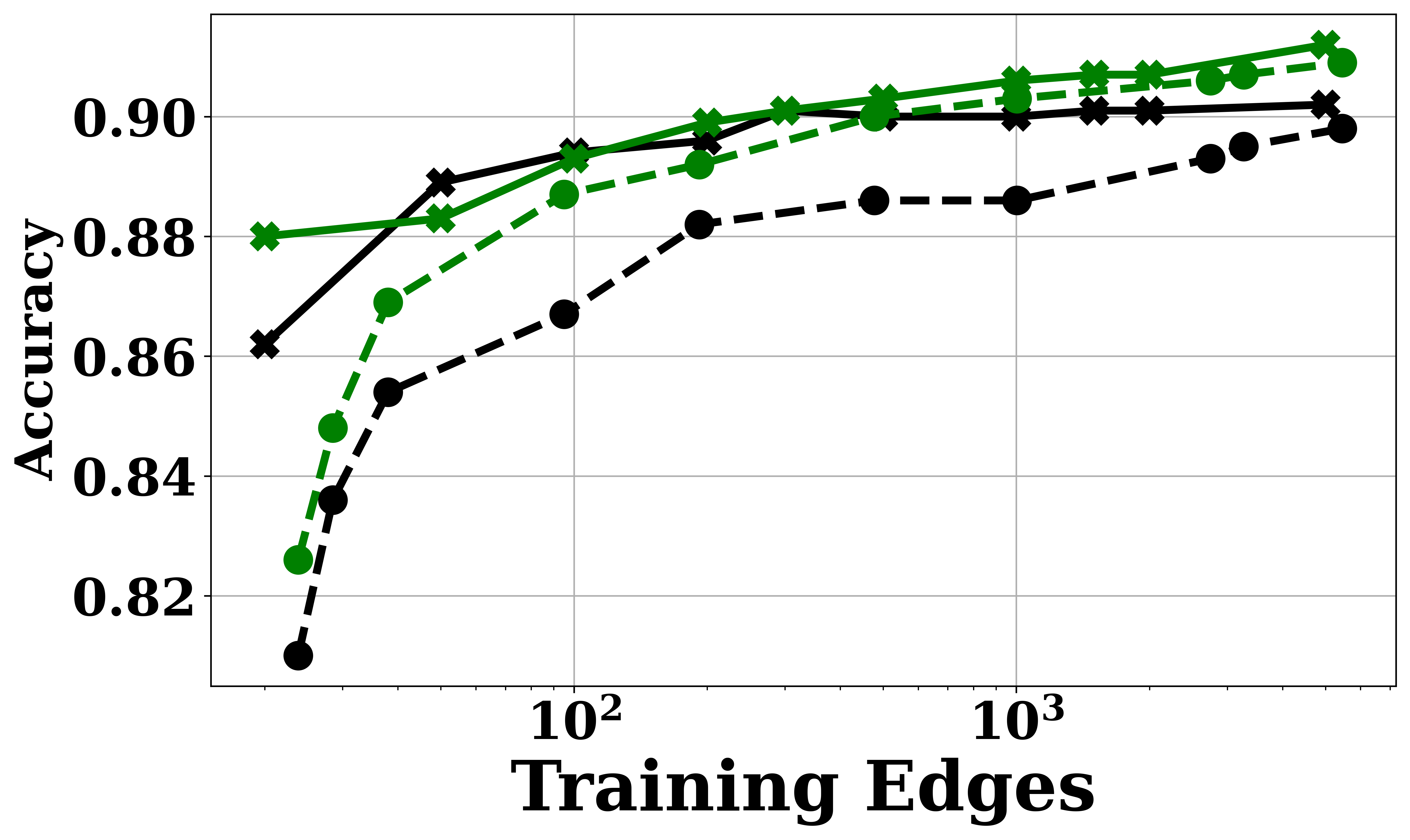}
        \vskip 0.25ex\caption{CollegeMSG}
        \label{fig:msg_labels}
    \end{subfigure}%
    \hfill
    \begin{subfigure}{0.66\columnwidth}
      \centering
        \includegraphics[width=0.8\linewidth]{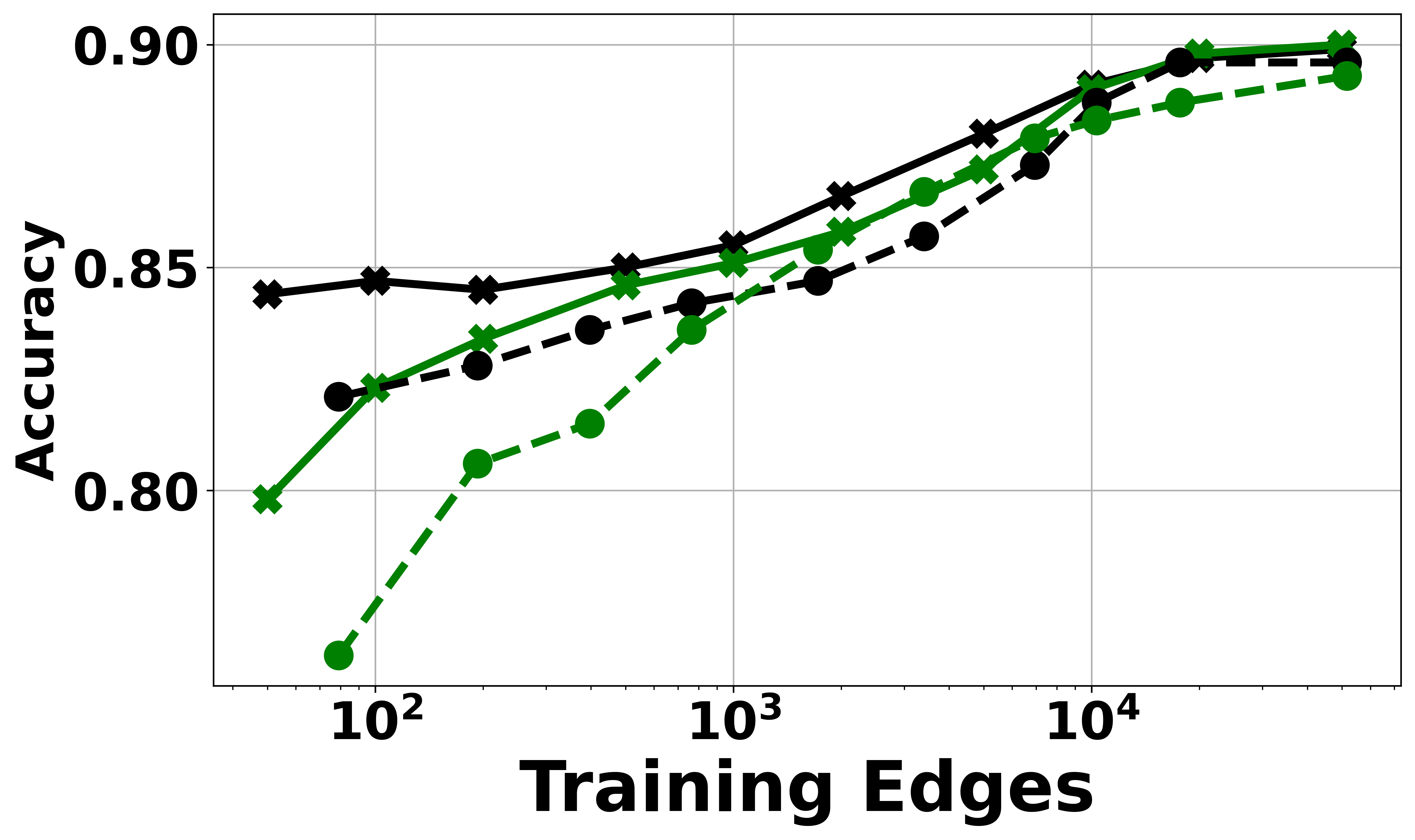}
        \vskip 0.25ex\caption{Twitter}
        \label{fig:tw_labels}
    \end{subfigure}%
    \vskip -1.5ex
    }
    \caption{Average prediction accuracy of MLP w/ $\mathbf{E}$ and GCN over all the definitions in the random and node-centric weakly supervised setting across different training sample sizes and datasets across 10 runs.} \label{fig:label_size}
     \vskip -2ex
\end{figure*}
    
\subsection{Additional Investigation}\label{sec:further}

In this section, we conducted further experiments to discuss not only the concerns drawn from the experiments in Section~\ref{Sec:emp_weak1} to Section~\ref{Sec:emp_un} but also show the insights drawn from existing literature that is related to model performance.

\begin{itemize}[leftmargin=*]
\item \textbf{Imbalance Issue:}  
As shown in Section~\ref{Sec:emp_weak1} and ~\ref{Sec:emp_weak2}, extreme class imbalances exist, such as CollegeMSG (Definition IV) and Twitter (Definition III). To address this, we apply two techniques: quantity reweighting and balanced hub node selection.  

Quantity reweighting~\cite{huang2016learning} is a common technique that assigns higher weights to minority classes during training. Specifically, for an edge $e_{ij} \in \mathcal{E}^{Tr}$ in class $k$, the assigned weight is:  
\begin{equation}\label{eq-qr}
\small 
    w_{ij}^{q} = \frac{|\mathcal{E}^{\text{Tr}}|}{|\{\mathcal{E}_k^{\text{Tr}}\}_{k = 1}^{C}|}
\end{equation}

Moreover, in the node-centric setting, we select hub nodes with balanced local class distributions to reduce bias.  

\hspace{2ex} The results in Tables~\ref{table:qw1} and~\ref{table:ave_qw} show that while these techniques improve performance in some cases (e.g., MLP w/ $\mathbf{E}$ for CollegeMSG IV and Twitter III in terms of prediction accuracy), they fail to significantly enhance overall accuracy. Macro-F1 scores remain low, indicating that imbalance remains challenging. 

\hspace{2ex}  Recent studies on imbalanced node/graph classification~\cite{chen2021topology, song2022tam, wang2022imbalanced} highlight the impact of topological imbalances on performance. We hypothesize that tie strength prediction also suffers from similar challenges. Inspired by recent imbalance edge classification work~\cite{cheng2024edge}, we believe addressing imbalance from a local topological perspective could lead to more effective solutions.

\vspace{0.5ex}

\begin{table}[t]
\centering 
\caption{Apply quantity imbalance techniques in the random and node-centric weakly supervised setting. 
}\label{table:qw1}
\vskip -0.75ex
 \footnotesize

\begin{tabular}{c|cc|cc}
\cline{2-5}
          & \multicolumn{2}{c|}{CollegeMSG IV} & \multicolumn{2}{c}{Twitter III} \\ \hline
Methods   & Acc           & Macro-F1           & Acc          & Macro-F1         \\ \hline
\multicolumn{5}{c}{Quantity reweight I} \\ \hline
MLP w/o $\mathbf{E}$ & 0.853  & 0.482   & 0.959  & 0.492   \\
MLP w/ $\mathbf{E}$  & 0.898  & 0.788   & 0.971  & 0.718   \\
GCN       & 0.908  & 0.761   & 0.954  & 0.504   \\
GTN       & 0.838  & 0.502   & 0.899  & 0.503   \\ \hline
\multicolumn{5}{c}{Quantity reweight II} \\ \hline
MLP w/o $\mathbf{E}$ & 0.878  & 0.479   & 0.935  & 0.492   \\
MLP w/ $\mathbf{E}$  & 0.820  & 0.709   & 0.967  & 0.640   \\
GCN       & 0.859  & 0.749   & 0.960  & 0.491   \\
GTN       & 0.893  & 0.473   & 0.915  & 0.493   \\ \hline
\multicolumn{5}{c}{Balanced hub nodes} \\ \hline
MLP w/o $\mathbf{E}$ & 0.870  & 0.490   & 0.953  & 0.493   \\
MLP w/ $\mathbf{E}$  & 0.908  & 0.798   & 0.964  & 0.581   \\
GCN       & 0.887  & 0.777   & 0.959  & 0.492   \\
GTN       & 0.825  & 0.503   & 0.949  & 0.496   \\ \hline
\end{tabular}
\vskip -1.25ex
\end{table}

\begin{table}[t]
\footnotesize 
\centering
\caption{Average weight differences between strong and weak ties of MLP w/ $\mathbf{E}$ after applying quantity imbalance techniques is significantly improved.}\label{table:ave_qw}
\vskip -0.75ex
\begin{tabular}{c|cc}
\cline{2-3}
\multicolumn{1}{c|}{} & CollegeMSG IV & Twitter III \\ \hline
Ground Truth          &     14.048          &      3.143       \\ \hline
Quantity reweight I   &      12.824        &      2.622       \\
Quantity reweight II  &       9.587        &      4.142       \\
Balanced Hub node     &       14.106        &      3.169       \\ \hline
\end{tabular}
\vskip -1.25ex
\end{table}

\item \textbf{Predicting Tie Strength in a Dynamic Setting:}  
Building on ~\cite{huang2018will}, which shows that tie strength evolves, we suggest that today's observed tie strength may not reflect its future state. This implies that predictions based on current data might forecast future tie strength rather than its present value. To test this, we used three methods (STC-Greedy, MLP w/ $\mathbf{E}$, and GCN) to predict tie strength in the initial Twitter network, then compared the results to the pseudo ground-truth after $\Delta T$ weeks. As shown in Figure~\ref{fig:pred_future}, how accuracy changes over time depends on the definition used. Definition VII improves as $\Delta T$ increases, suggesting strong friendships take time to form. In contrast, Definitions II and III decline, indicating their predictions reflect current tie strength rather than future development.

\item \textbf{Impact of Experimental Settings:}  
Based on the performance shown in Figure~\ref{fig:settingvs} and in Appendix, overall, existing methods perform similarly in both weakly supervised settings, with the performance on the random setting being slightly better. To explore how models perform under different weakly supervised settings, we varied the training sample size. Figure~\ref{fig:label_size} shows the average accuracy of MLP w/ $\mathbf{E}$ and GCN across datasets over 10 runs. Models generally perform better in the random setting than in the node-centric setting when the training size is small, likely because the latter introduces bias based on hub nodes. However, as training size increases, this gap narrows, suggesting that a broader variety of edges reduces initial bias.

\end{itemize}

\vspace{-2ex}

\section{Limitations and Future Directions}~\label{Sec:lim}
Since this work presents a comprehensive benchmark for tie strength prediction, our efforts were not to develop the most state-of-the-art prediction method. However, we offer key insights into future advances based on our experimental findings. Overall, neural network models outperform heuristic methods. This is likely because neural network models are parametrized and can learn the complicated patterns in the data that are slipping through heuristic methods. When the training sample size increases, neural network models that can effectively leverage the information in local neighborhoods are more likely to perform better in different scenarios. Besides, semantic features depict the potential to improve performance, but when incorporating them in GNNs, the aggregating and updating mechanism needs careful design. However, despite relatively good performance in weakly supervised settings, neural networks can not be directly implemented for unsupervised tie strength prediction tasks, as no training labels are provided.
    
Also, we can conclude that even though heuristic models with predefined rules might outperform other baselines in biased training scenarios, they generally do not perform well in most other situations. This inconsistency demonstrates that the effectiveness of predefined rules is not guaranteed. For STC-Greedy, as it is not a data-driven learning method, it is hard to perform well in complex scenarios. Moreover, none of the existing models can distinguish strong and weak ties to reflect the weight differences effectively. 

Based on the above, it would be worthwhile to focus on (1) developing better mechanisms to integrate semantic features into GNN models, (2) extending methods to enable neural networks to perform well even with imbalanced training data~\cite{cheng2025edge}, i.e., either the inherent label quantity imbalance or structural from the node-centric setting, and (3) designing graph contrastive learning models~\cite{you2020graph,wang2022graph} to better leverage unlabeled data that might help distinguish between large and small weight edges for weakly/unsupervised settings.

\section{Conclusion}\label{sec:con}

In this study, we address the challenge in current tie strength prediction research by introducing \textbf{BTS}, a comprehensive \textbf{B}enchmark for \textbf{T}ie \textbf{S}trength prediction. Specifically, we categorize mainstream understandings of tie strength into seven standardized definitions and validate their effectiveness by analyzing class distributions and correlations across these definitions on three well-curated real-world datasets. Then we propose a standardized framework to evaluate the quality of pseudo-labels from the perspective of tie resilience. To benchmark tie strength prediction, we evaluate the performance of representative models, and explore 
the effects of different experimental settings, models, and evaluation criteria on prediction performance. Overall, \textbf{BTS} establishes a standardized foundation for evaluating and advancing tie strength prediction methodologies.

\vspace{-1ex}
\section{Acknowledgments} This research is supported by the National Science Foundation under grant number IIS2239881.

\bibliographystyle{ACM-Reference-Format}
\balance 
\bibliography{reference}

\appendix
\vspace{-0.75em}
\section{Appendix}\label{appendix}

Below, we present all additional experimental results, including prediction accuracy and tie weight differences for both weakly supervised settings, as well as STC-Greedy performance under the unsupervised setting.

\vspace{1em}

\begin{table}[H]
\vspace{-1.25em}
\scriptsize 
\centering 
\setlength\tabcolsep{1.5pt}
\caption{
Tie weight difference (random weakly setting).}\label{table:ave_ba1}
\vspace{-1.5em}
\begin{tabular}{c|ccccccc|c}
\hline
Method       & I     & II     & III   & IV    & V     & VI    & VII   & Avg.   \\ \hline
\multicolumn{9}{c}{Bitcoin\_alpha} \\ \hline
Ground Truth & 1.658 & 6.215  & 6.227 & 5.872 & 2.966 & 3.960 & 3.985 & 4.698 \\ \hline
Heuristic    & -0.04 & 0.791  & 0.521 & 0.958 & -0.045 & 0.275 & 0.123 & 0.368 \\
MLP w/o $\mathbf{E}$    & 1.200 & -3.489 & -3.489 & -3.489 & -0.027 & -3.489 & -3.489 & -2.467 \\
MLP w/ $\mathbf{E}$     & 0.212 & -0.578 & -0.578 & -1.432 & 0.568  & -1.856 & -0.349 & -0.573 \\
GCN          & 1.107 & -3.488 & -3.488 & -3.488 & 0.240  & -1.156 & -1.247 & -1.788 \\
GTN          & -0.011 & 0.314 & 0.458  & 0.148  & 0.089  & -0.259 & -0.337 & 0.058  \\ \hline
Avg.          & 0.688 & -0.872 & -0.892 & -0.905 & 0.399  & -0.839 & -0.902 & -0.332 \\ \hline
\multicolumn{9}{c}{CollegeMSG} \\ \hline
Ground Truth & 5.323 & 10.253 & 10.392 & 14.048 & 6.699 & 10.217 & 11.132 & 9.866 \\ \hline
Heuristic    & 0.461 & 1.737  & 2.014  & 0.576  & -0.316 & -1.511 & -1.610 & 0.193  \\
MLP w/o $\mathbf{E}$    & 0.177 & -4.311 & -4.311 & -3.022 & -1.425 & -3.018 & -4.311 & -2.889 \\
MLP w/ $\mathbf{E}$     & 5.360 & 10.268 & 10.449 & 10.279 & 9.594 & 12.322 & 13.166 & 10.777 \\
GCN          & 6.009 & 10.191 & 11.764 & 2.634  & 9.403 & 7.061  & 13.052 & 8.302  \\
GTN          & 1.664 & -2.606 & -4.320 & -4.320 & -0.953 & -4.320 & -4.320 & -2.679 \\ \hline
Avg.          & 3.832 & 3.755  & 4.665  & 3.699  & 3.667  & 3.392  & 4.185  & 3.885 \\
\hline

\multicolumn{9}{c}{Twitter} \\ \hline
Ground Truth & 0.458 & 2.499  & 3.143  & 0.532  & 0.641  & 0.100  & 0.635  & 1.287  \\ \hline
Heuristic    & 0.125 & 0.411  & 0.739  & 0.078  & 0.064  & -0.287 & -0.064 & 0.152  \\
MLP w/o $\mathbf{E}$    & -0.071 & -0.088 & -0.101 & -0.078 & -0.036 & 0.002  & -0.063 & -0.062 \\
MLP w/ $\mathbf{E}$     & 0.124 & 0.478  & -0.173 & -0.173 & 0.642  & 0.086  & -0.173 & 0.116  \\
GCN          & 0.133 & -0.173 & -0.173 & -0.173 & 0.144  & -0.171 & -0.173 & -0.083 \\
GTN          & -0.018 & -0.173 & -0.173 & -0.173 & -0.008 & -0.054 & -0.068 & -0.095 \\ \hline
Avg.          & 0.125 & 0.476  & 0.543  & 0.151  & 0.241  & -0.054 & 0.016  & 0.214 
\\
\hline
\end{tabular}
\end{table}

\begin{table*}[t]
\footnotesize 
\setlength\tabcolsep{3pt}
\setlength{\extrarowheight}{0.6pt}
\caption{Prediction accuracy of existing models on three datasets across seven definitions in the random weakly supervised setting. Overall, models perform worse in Definition I, V, and VII, with MLP w/ $\mathbf{E}$ and GCN performing best.}
\label{table:res1}
\begin{tabular}{c|c|ccc|c|cccc}
\hline
\textbf{Definitions} & \textbf{Model} & \textbf{Bitcoin\_alpha} & \textbf{CollegeMSG} & \textbf{Twitter} & \multicolumn{1}{c|}{\textbf{Definitions}} & \multicolumn{1}{c|}{\textbf{Model}} & \textbf{Bitcoin\_alpha} & \textbf{CollegeMSG} & \textbf{Twitter} \\\hline
\multirow{5}{*}{\shortstack{I\\(Avg. 0.720)}} & Heuristic & 0.695 & 0.572 & 0.739 & \multicolumn{1}{c|}{\multirow{5}{*}{\shortstack{V\\(Avg. 0.636)}}} & \multicolumn{1}{c|}{Heuristic} & 0.588 & 0.582 & 0.648 \\
& MLP w/o $\mathbf{E}$ & 0.653 & 0.532 & 0.756 & & \multicolumn{1}{c|}{MLP w/o $\mathbf{E}$} & 0.521 & 0.610 & 0.731 \\
& MLP w/ $\mathbf{E}$ & 0.789 & 0.845 & 0.909 & & \multicolumn{1}{c|}{MLP w/ $\mathbf{E}$} & 0.560 & 0.845 & 0.743 \\
& GCN & 0.719 & 0.824 & 0.885 & & \multicolumn{1}{c|}{GCN} & 0.537 & 0.809 & 0.713 \\
& GTN & 0.640 & 0.552 & 0.695 & & \multicolumn{1}{c|}{GTN} & 0.507 & 0.517 & 0.624 \\\hline
\multirow{5}{*}{\shortstack{II\\(Avg. 0.852)}} & Heuristic & 0.749 & 0.737 & 0.910 & \multicolumn{1}{c|}{\multirow{5}{*}{\shortstack{VI\\(Avg. 0.713)}}} & \multicolumn{1}{c|}{Heuristic} & 0.679 & 0.751 & 0.827 \\
& MLP w/o $\mathbf{E}$ & 0.794 & 0.758 & 0.930 & & \multicolumn{1}{c|}{MLP w/o $\mathbf{E}$} & 0.741 & 0.781 & 0.573 \\
& MLP w/ $\mathbf{E}$ & 0.783 & 1.000 & 0.939 & & \multicolumn{1}{c|}{MLP w/ $\mathbf{E}$} & 0.726 & 0.900 & 0.544 \\
& GCN & 0.794 & 0.976 & 0.931 & & \multicolumn{1}{c|}{GCN} & 0.674 & 0.863 & 0.637 \\
& GTN & 0.789 & 0.757 & 0.931 & & \multicolumn{1}{c|}{GTN} & 0.650 & 0.794 & 0.558 \\\hline
\multirow{5}{*}{\shortstack{III\\(Avg. 0.874)}} & Heuristic & 0.756 & 0.773 & 0.960 & \multicolumn{1}{c|}{\multirow{5}{*}{\shortstack{VII\\(Avg. 0.811)}}} & \multicolumn{1}{c|}{Heuristic} & 0.692 & 0.783 & 0.865 \\
& MLP w/o $\mathbf{E}$ & 0.828 & 0.789 & 0.865 & & \multicolumn{1}{c|}{MLP w/o $\mathbf{E}$} & 0.766 & 0.830 & 0.865 \\
& MLP w/ $\mathbf{E}$ & 0.816 & 0.965 & 0.867 & & \multicolumn{1}{c|}{MLP w/ $\mathbf{E}$} & 0.756 & 0.907 & 0.867 \\
& GCN & 0.828 & 0.939 & 0.960 & & \multicolumn{1}{c|}{GCN} & 0.731 & 0.907 & 0.867 \\
& GTN & 0.825 & 0.789 & 0.960 & & \multicolumn{1}{c|}{GTN} & 0.666 & 0.830 & 0.841 \\\hline
\multirow{5}{*}{\shortstack{IV\\(Avg. 0.872)}} & Heuristic & 0.879 & 0.873 & 0.829 & \multicolumn{1}{c|}{\multirow{5}{*}{Avg}} & \multicolumn{1}{c|}{Heuristic} & 0.720 & 0.724 & 0.825 \\
& MLP w/o $\mathbf{E}$ & 0.889 & 0.882 & 0.843 & & \multicolumn{1}{c|}{MLP w/o $\mathbf{E}$} & 0.742 & 0.740 & 0.808 \\
& MLP w/ $\mathbf{E}$ & 0.882 & 0.890 & 0.845 & & \multicolumn{1}{c|}{MLP w/ $\mathbf{E}$} & 0.759 & 0.907 & 0.830 \\
& GCN & 0.889 & 0.905 & 0.845 & & \multicolumn{1}{c|}{GCN} & 0.739 & 0.889 & 0.834 \\
& GTN & 0.886 & 0.894 & 0.845 & & \multicolumn{1}{c|}{GTN} & 0.709 & 0.733 & 0.779 \\\hline
\end{tabular}
\end{table*}

\begin{table*}[h]
\footnotesize 
\setlength\tabcolsep{3pt}
\setlength{\extrarowheight}{0.6pt}
\caption{Prediction accuracy of existing models on three datasets across seven definitions in the node-centric weakly supervised setting. Similarly, overall models perform worse in Definition I, V, and VII, and MLP w/ $\mathbf{E}$ and GCN perform best. 
}\label{table:res2}
\begin{tabular}{c|c|ccc||ccccc}
\hline
\textbf{Definitions} &
  \textbf{Model} &
  \textbf{Bitcoin\_ahpha} &
  \textbf{CollegeMSG} &
  \textbf{Twitter} &
  \multicolumn{1}{c|}{\textbf{Definitions}} &
  \multicolumn{1}{c|}{\textbf{Model}} &
  \textbf{Bitcoin\_ahpha} &
  \textbf{CollegeMSG} &
  \textbf{Twitter} \\ \hline
\multirow{5}{*}{\shortstack{I\\ \\(Avg. 0.715)}}   & Heuristic  & 0.707 & 0.559 & 0.754 & \multicolumn{1}{c|}{\multirow{5}{*}{\shortstack{V\\ \\(Avg. 0.631)}}}   & \multicolumn{1}{c|}{Heuristic}  & 0.509 & 0.563 & 0.597 \\
                     & MLP w/o $\mathbf{E}$  & 0.746 & 0.506 & 0.663 & \multicolumn{1}{c|}{}                     & \multicolumn{1}{c|}{MLP w/o $\mathbf{E}$}  & 0.526 & 0.638 & 0.735 \\
                     & MLP w/ $\mathbf{E}$   & 0.833 & 0.817 & 0.914 & \multicolumn{1}{c|}{}                     & \multicolumn{1}{c|}{MLP w/ $\mathbf{E}$}   & 0.533 & 0.813 & 0.736 \\
                     & GCN        & 0.741 & 0.817 & 0.757 & \multicolumn{1}{c|}{}                     & \multicolumn{1}{c|}{GCN}        & 0.516 & 0.798 & 0.733 \\
                     & GTN        & 0.650 & 0.511 & 0.743 & \multicolumn{1}{c|}{}                     & \multicolumn{1}{c|}{GTN}        & 0.495 & 0.588 & 0.685 \\ \hline
\multirow{5}{*}{\shortstack{II\\ \\(Avg. 0.841)}}  & Heuristic  & 0.790 & 0.712 & 0.930 & \multicolumn{1}{c|}{\multirow{5}{*}{\shortstack{VI\\ \\(Avg. 0.700)}}}  & \multicolumn{1}{c|}{Heuristic}  & 0.712 & 0.706 & 0.812 \\
                     & MLP w/o $\mathbf{E}$  & 0.794 & 0.757 & 0.931 & \multicolumn{1}{c|}{}                     & \multicolumn{1}{c|}{MLP w/o $\mathbf{E}$}  & 0.741 & 0.794 & 0.454 \\
                     & MLP w/ $\mathbf{E}$   & 0.794 & 0.918 & 0.931 & \multicolumn{1}{c|}{}                     & \multicolumn{1}{c|}{MLP w/ $\mathbf{E}$}   & 0.729 & 0.877 & 0.458 \\
                     & GCN        & 0.794 & 0.884 & 0.931 & \multicolumn{1}{c|}{}                     & \multicolumn{1}{c|}{GCN}        & 0.741 & 0.855 & 0.614 \\
                     & GTN        & 0.760 & 0.758 & 0.931 & \multicolumn{1}{c|}{}                     & \multicolumn{1}{c|}{GTN}        & 0.712 & 0.776 & 0.524 \\ \hline
\multirow{5}{*}{\shortstack{III\\ \\(Avg. 0.873)}} & Heuristic  & 0.821 & 0.751 & 0.960 & \multicolumn{1}{c|}{\multirow{5}{*}{\shortstack{VII\\ \\(Avg. 0.812)}}} & \multicolumn{1}{c|}{Heuristic}  & 0.724 & 0.745 & 0.867 \\
                     & MLP w/o $\mathbf{E}$  & 0.828 & 0.789 & 0.960 & \multicolumn{1}{c|}{}                     & \multicolumn{1}{c|}{MLP w/o $\mathbf{E}$}  & 0.766 & 0.829 & 0.867 \\
                     & MLP w/ $\mathbf{E}$   & 0.828 & 0.969 & 0.960 & \multicolumn{1}{c|}{}                     & \multicolumn{1}{c|}{MLP w/ $\mathbf{E}$}   & 0.766 & 0.883 & 0.867 \\
                     & GCN        & 0.828 & 0.885 & 0.960 & \multicolumn{1}{c|}{}                     & \multicolumn{1}{c|}{GCN}        & 0.766 & 0.832 & 0.853 \\
                     & GTN        & 0.805 & 0.789 & 0.960 & \multicolumn{1}{c|}{}                     & \multicolumn{1}{c|}{GTN}        & 0.736 & 0.830 & 0.853 \\ \hline
\multirow{5}{*}{\shortstack{IV\\ \\(Avg. 0.867)}}  & Heuristic  & 0.861 & 0.881 & 0.799 & \multicolumn{1}{c|}{\multirow{5}{*}{Avg}} &   \multicolumn{1}{c|}{Heuristic}  &   0.732 & 0.702 & 0.817                                \\
                     & MLP w/o $\mathbf{E}$  & 0.889 & 0.894 & 0.845 &  \multicolumn{1}{c|}{}                     & \multicolumn{1}{c|}{MLP w/o $\mathbf{E}$}  &    0.756 & 0.744 & 0.779                                                         \\
                     & MLP w/ $\mathbf{E}$   & 0.889 & 0.894 & 0.845 & \multicolumn{1}{c|}{}                     & \multicolumn{1}{c|}{MLP w/ $\mathbf{E}$}  &   0.767 &0.882& 0.816                                                           \\
                     & GCN        & 0.889 & 0.894 & 0.845 & \multicolumn{1}{c|}{}                     & \multicolumn{1}{c|}{GCN} &     0.754 & 0.852 & 0.813                                                         \\
                     & GTN        & 0.882 & 0.894 & 0.803 & \multicolumn{1}{c|}{}                     & \multicolumn{1}{c|}{GTN} &  0.720 & 0.735 & 0.786                                                            \\ \hline
\end{tabular}
\end{table*}

\begin{table}[H]
\scriptsize 
\centering 
\setlength\tabcolsep{1.5pt}
\caption{
Tie weight difference (node-centric setting). 
}\label{table:ave_ba2}
\vspace{-1.5em}
\begin{tabular}{c|ccccccc|c}
\hline
Method       & I     & II     & III   & IV    & V     & VI    & VII   & Avg.   \\ \hline
\multicolumn{9}{c}{Bitcoin\_alpha} \\ \hline
Ground Truth & 1.657 & 6.217  & 6.229 & 5.870 & 2.967 & 3.961 & 3.984 & 4.699 \\ \hline
Heuristic    & -0.376 & 0.796 & 1.282 & 0.139 & 0.115 & -0.101 & -0.054 & 0.257 \\
MLP w/o $\mathbf{E}$    & 3.487 & -3.487 & -3.487 & -3.487 & 0.009 & -3.487 & -3.487 & -1.991 \\
MLP w/ $\mathbf{E}$     & 0.595 & -3.489 & -3.489 & -3.489 & 0.312 & 0.211  & -3.489 & -1.263 \\
GCN          & 2.153 & -3.489 & -3.489 & -3.489 & 0.011 & -3.489 & -3.489 & -1.897 \\
GTN          & 1.056 & 0.186  & 0.261  & -0.895 & -0.057 & -0.430 & -0.489 & -0.052 \\ \hline
Avg.          & 1.429 & -0.878 & -0.948 & -1.895 & 0.560 & -0.889 & -1.146 & -0.481 \\ \hline
\multicolumn{9}{c}{CollegeMSG} \\ \hline
Ground Truth & 5.354 & 10.316 & 10.441 & 14.137 & 6.744 & 10.259 & 11.216 & 9.924 \\ \hline
Heuristic    & 1.017 & 0.825  & 4.142  & -0.405 & -0.487 & -0.163 & 0.786  & 0.816 \\
MLP w/o $\mathbf{E}$    & 0.182 & -4.331 & -4.331 & -4.331 & -4.331 & -4.331 & -4.331 & -3.029 \\
MLP w/ $\mathbf{E}$     & 6.500 & 5.438  & 10.313 & -4.331 & 8.458  & 11.558 & 7.218  & 6.165 \\
GCN          & 6.166 & 6.087  & 5.236  & -4.330 & 7.667  & 10.625 & 0.005  & 4.208 \\
GTN          & -1.191 & -0.780 & -4.324 & -4.324 & -1.309 & -3.226 & -4.324 & -2.783 \\ \hline
Avg.          & 2.338 & 2.259  & 3.596  & -1.764 & 2.957  & 3.453  & 1.381  & 2.460  \\ \hline
\multicolumn{9}{c}{Twitter} \\ \hline
Ground Truth & 0.458 & 2.499  & 3.143  & 0.531  & 0.641  & 0.100  & 0.635  & 1.287  \\ \hline
Heuristic    & 0.544 & 0.815  & 0.253  & 0.081  & 0.089  & -0.298 & -0.106 & 0.197  \\
MLP w/o $\mathbf{E}$    & -0.049 & -0.173 & -0.173 & -0.173 & -0.173 & -0.119 & -0.173 & -0.147 \\
MLP w/ $\mathbf{E}$     & 0.209  & -0.173 & -0.173 & -0.173 & 0.839  & -0.269 & -0.173 & 0.012  \\
GCN          & -0.173 & -0.173 & -0.173 & -0.173 & 0.130  & -0.029 & -0.173 & -0.109 \\
GTN          & -0.050 & -0.163 & -0.163 & -0.119 & -0.033 & 0.050  & -0.101 & -0.085 \\ \hline
Avg.          & 0.157  & 0.438  & 0.452  & -0.004 & 0.249  & -0.094 & -0.015 & 0.169  
\\ 
\hline
\end{tabular}
\end{table}

\begin{table}[H]
\small
\setlength\tabcolsep{3pt}
\caption{Accuracy of STC-Greedy in unsupervised setting. 
}\label{table:res3}
\begin{tabular}{c|ccccccc|c}
\hline
Dataset        & I       & II      & III     & IV      & V       & VI      & VII     & Avg. \\ \hline
Bitcoin\_alpha & 0.295   & 0.742   & 0.769   & 0.819   & 0.540   & 0.697   & 0.718   & 0.654   \\  
CollegeMSG     & 0.529   & 0.721   & 0.751   & 0.844   & 0.616   & 0.756   & 0.787   & 0.715   \\
Twitter        & 0.736   & 0.898   & 0.926   & 0.816   & 0.713   & 0.406   & 0.839   & 0.762   \\ \hline
Avg.            & 0.520       &    0.787     &   0.815      &   0.826      &   0.623      &  0.620       &  0.781       &  0.710   \\ \hline
\end{tabular}
\vskip 1ex
\end{table}

\end{document}